\documentclass[lettersize, journal]{IEEEtran}		

\usepackage[T1]{fontenc}
\usepackage{amsmath,amsfonts}
\usepackage{algorithmic}
\usepackage{adjustbox}
\usepackage{caption}

\usepackage{subcaption}
\usepackage{array}
\usepackage[caption=false,font=normalsize,labelfont=sf,textfont=sf]{subfig}
\usepackage{textcomp}
\usepackage{stfloats}
\usepackage{url}
\usepackage{authblk}

\usepackage{verbatim}
\usepackage{graphicx}
\usepackage{bm} 

\usepackage[table]{xcolor} 
\usepackage{tabularray}
\usepackage{tabularx}

\usepackage{booktabs}
\usepackage{makecell}

\usepackage{multicol}
\usepackage{multirow}
\usepackage{balance}	

\usepackage[switch]{lineno} 

\usepackage[
  backend=biber,
  style=numeric-comp,   
  sorting=none
]{biblatex}

\graphicspath{ {./Figures/} }
\def\BibTeX{{\rm B\kern-.05em{\sc i\kern-.025em b}\kern-.08em
    T\kern-.1667em\lower.7ex\hbox{E}\kern-.125emX}}

\newcolumntype{s}{>{\hsize=.33\hsize\linewidth=\hsize}X}
\newcolumntype{D}{>{\hsize=.4\hsize\linewidth=\hsize}X} 

\usepackage{ragged2e}   
\usepackage{amssymb}

\newcolumntype{L}[1]{>{\RaggedRight\arraybackslash}p{#1}}

\begin{document}

\title{Spatially multiplexed opto-electronic Ising machine with single-shot parallel readout of 100,000 spins

\thanks{\hrule \vspace{0.25em}  *Correspondence and requests should be addressed to T.S. (email: toon.jan.b.sevenants@vub.be) or to G.V. (email: guy.verschaffelt@vub.be)}}

\author[*1, 2]{Toon Sevenants}
\author[1]{Joren Van Cauwenberge}
\author[2]{Kaijun Wu}
\author[2]{Peter Bienstman}
\author[1]{Guy Van der Sande}
\author[1]{Guy Verschaffelt}

\affil[1]{Applied Physics Research Group, Vrije Universiteit Brussel, Brussels, Belgium} 
\affil[2]{Photonics Research Group, Ghent University-imec, Gent, Belgium}

\maketitle

\begin{abstract}
Ising machines offer a promising route to solving NP-hard combinatorial optimization problems by mapping them onto the ground state of the Ising model. Free-space optical implementations are attractive for their inherent parallelism, scalability, and energy efficiency, yet nearly all rely on individual, sequential spin updates, leaving most of this parallelism unexploited. Here, we reclaim it with a free-space optical analog Ising machine built from off-the-shelf components in which every spin is addressed in parallel and read out in a single camera shot, while letting the system evolve under analog dynamics rather than being written and read one spin at a time. We demonstrate this architecture with over 100,000 spins, the largest spatially multiplexed analog Ising machine reported to date, with a realistic path beyond one million spins limited only by spatial-light-modulator and camera resolution. Across sparse and dense MaxCut and fully connected number-partitioning instances, our machine matches state-of-the-art solution quality. Because every spin updates in parallel, throughput increases with system size, unlike time-multiplexed machines, which slow as problems grow.  The same free-space geometry also lets us solve more than 50 problem instances simultaneously, further boosting throughput. Together, these results mark an important step towards a scalable, and energy-efficient Ising machine.
\end{abstract}

\section*{}
\label{sec: Introduction}
\IEEEPARstart{G}{lobal} data processing now consumes a rapidly growing share of the world's energy, and digital computation is approaching physical limits on efficiency as transistor scaling slows down ~\cite{chen2025data, shehabi20242024, Shalf2020MooresLaw}. Both trends have intensified the search for computing paradigms beyond the digital von Neumann architecture. Physical implementations of the Ising model, called Ising machines (IMs), are among the most promising alternatives. They operate by mapping the cost function of an optimization problem onto a network of interacting binary spins and use collective physical dynamics, rather than sequential logic, to search for low-energy configurations of that spin network ~\cite{mohseni2022IMhardwaresolver, kalinin2022Computational}.

Many real-world optimization problems across science and industry, including protein folding, job-shop scheduling, portfolio optimization, prime factorization, vehicle routing, and wireless resource allocation in 5G networks, are NP-hard, which makes them inherently hard to solve on digital computers, but they can be reformulated within the Ising model framework ~\cite{lucas2014ising, buonaiuto2023best, inoue2021traffic, lin2025multi, jiang2018quantum, irback2022folding}. Fast, energy-efficient, and scalable IM implementations could therefore have broad impact across these domains.

The shift from sequential computation to parallel, physics-driven relaxation is what underlies the prospect of improved energy and time efficiency. However, these advantages only become significant at large scale, which is why scaling has played a central role in novel hardware implementations. Realizing this potential therefore requires IM platforms that combine large-scale spin coupling with programmable interactions, enabling the implementation of diverse problems while maintaining stable operation over long periods of time.

Several physical platforms have already been explored, including photonic, magnetic, acoustic, and electronic implementations ~\cite{bohm2019poor, 200GOPS_IM_2025, Zhenhua2024OIM, Cen2022CIM, honjo2021100, litvinenko2025acousticwave, sutton2017nanomagnets, jain2021Dwave, chen2024memristors, ahmed2021Ringoscillator, wu2025monolithically, veraldi2025SPIM}. However, most of the mature systems, including the widely adopted opto-electronic-oscillator Ising machine ~\cite{bohm2019poor, 200GOPS_IM_2025, Zhenhua2024OIM, Cen2022CIM} and coherent Ising machine ~\cite{honjo2021100}, represent every spin using a single reused nonlinear element, multiplexed in time. This lets a small set of physical components emulate a large spin network, but it forces system complexity, latency, and synchronization overhead to grow with problem size, since spins can only be updated sequentially. 

\begin{figure*}[t!]
    \centering
    \includegraphics[width=\textwidth]{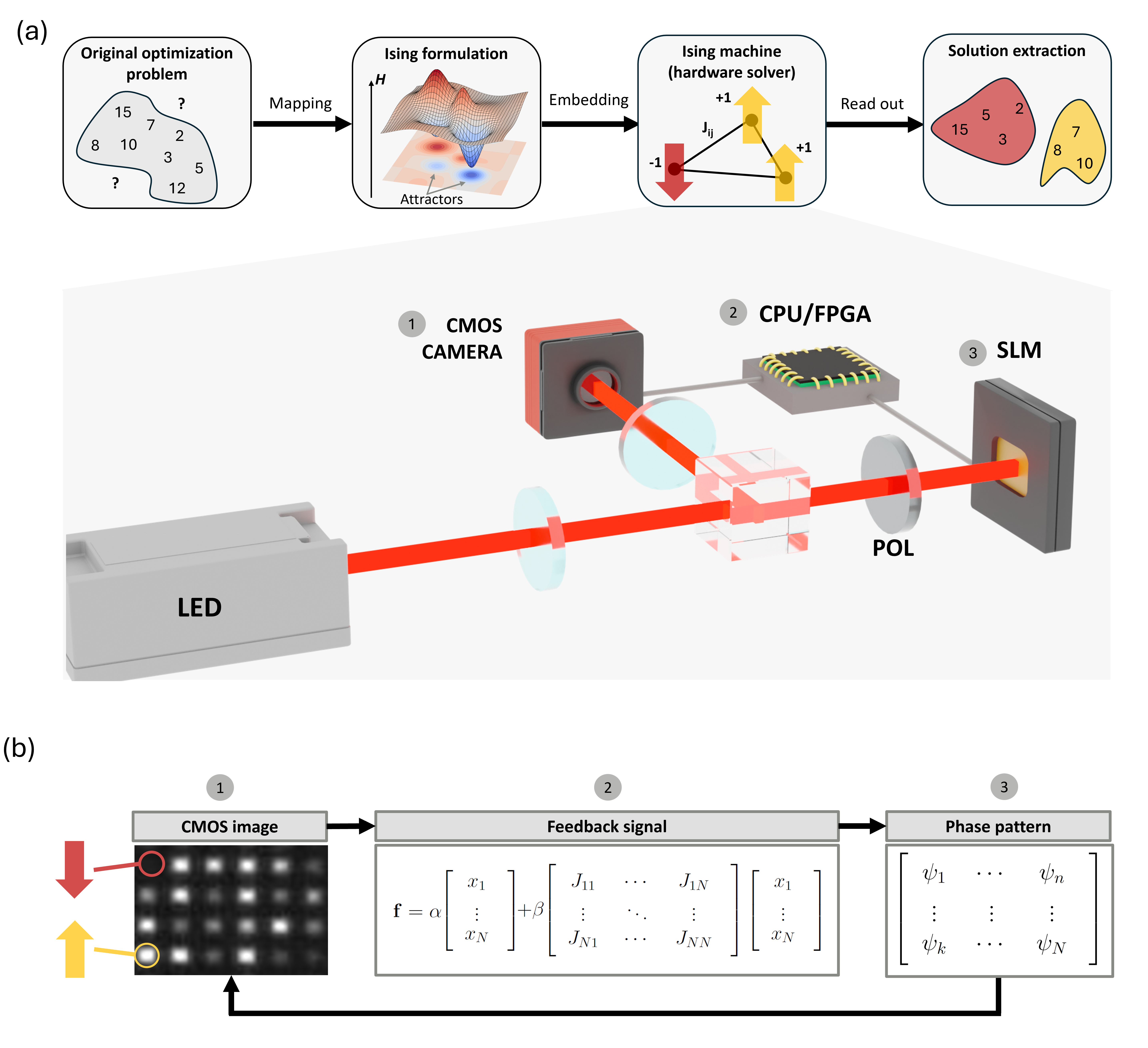}
    \caption{\textbf{Illustration of the proposed analog spatially-multiplexed Ising machine (ASIM) and its operational principle.}
\textbf{(a)} Combinatorial optimization problems can be solved using an analog Ising machine by mapping the cost function of the original problem (e.g., number partitioning) onto an Ising energy landscape. The problem variables are embedded into physical degrees of freedom, where each spin is represented by a measurable quantity (e.g., optical intensity). The Ising machine subsequently explores the Ising energy landscape by evolving its spin variables. At steady state, the associated binary spin configuration ($\{\sigma_i\}=\pm1$) is read out as an approximate solution. Our proposed ASIM implementation uses off-the-shelf components: an LED source, CMOS camera, digital processor (CPU/FPGA), spatial light modulator (SLM), and polarizer (POL).
\textbf{(b)} The iterative feedback loop comprises three steps: \textcircled{1} the CMOS camera captures the spin-state image, where bright and dark squares represent spin-up and spin-down, respectively. \textcircled{2} These analog intensities are processed digitally to compute the feedback signal. \textcircled{3} This signal is sent to the SLM, which encodes it as a phase pattern that, via the POL, modulates the intensity measured by the CMOS camera.}
    \label{fig:SLM_setup}
\end{figure*}

Spatial multiplexing offers a fundamentally different route, allowing all spins to evolve in parallel, which better exploits the intrinsic parallelism of analog photonics and removes the overhead of time-encoding entirely. In practice, however, spatial scaling has been difficult. Implementations based on parallel arrays of Mach-Zehnder modulators (MZMs) face significant obstacles in terms of footprint, power consumption, fabrication variability, and control complexity, making scaling to thousands of coupled spins nontrivial~\cite{Zhenhua2024OIM}. Free-space, spatial-light-modulator (SLM)-based IMs sidestep these fabrication constraints and have already reached large spin counts~\cite{veraldi2025SPIM, pierangeli2019SPIM, fang2021SPIM, ye2023SPIM, huang2021AntiferromagneticSPIM}. However, in these implementations the optics is used only to evaluate the Ising Hamiltonian for a given spin configuration, while the actual search for the ground state is carried out by an external heuristic algorithm, typically Metropolis--Hastings-style simulated annealing. A standard step in that heuristic proposes and evaluates a single spin flip, no matter how many spins the SLM addresses in parallel, so visiting every spin once already takes $N$ such steps. 
This entire process requires large computational assistance that can not be fully parallelized.  
The optical hardware thus acts as an accelerated cost-function evaluator rather than as the dynamical system doing the optimization. 

Systems that instead let the dynamics drive the spins toward the ground state can be fully parallelized and do not require an external search algorithm in the loop. However, such implementations have so far been confined to substantially smaller scales. For example, recent demonstrations of analog, spatially multiplexed IMs, including free-space and integrated photonic implementations, reach only on the order of $\sim 10$ spins~\cite{kalinin2025analog, wu2025monolithically}, as increasing the number of spins also increases the number of components that must be tightly integrated into a single fabricated system, highlighting the gap between conceptual scalability and practical realization.

Here, we demonstrate a new opto-electronic implementation of an analog spatially-multiplexed Ising machine (ASIM) that avoids these bottlenecks and goes well beyond the scale that current analog spatially-multiplexed IMs have reached, with a simple layout that is built from an SLM, a linear polarizer and a CMOS camera, in which all spins are read out in a single camera shot and updated simultaneously. A gradient-descent-based feedback signal drives the system toward low-energy spin configurations. We demonstrate this architecture at up to 110{,}889 spins, the largest spin count reported for a spatially multiplexed analog IM, while the proposed system offers a realistic path to go beyond $10^6$ spins set only by available SLM pixel count and CMOS resolution. This approach combines the scalability of free-space optics with the flexibility of programmable interactions, establishing a route toward large-scale analog optimization hardware that is structurally unavailable to time-multiplexed architectures.

The IM we present is a proof-of-concept, using general-purpose hardware. While the current implementation is not yet optimized, since it is constructed out of off-the-shelf components, it already shows favorable scaling with problem size that is fundamentally unavailable to time-multiplexed architectures, and will benefit directly from continued advances in SLM and CMOS-camera technology.

\section{Results}
\subsection{Operational principle}
To solve a problem with an IM, the target optimization task is first mapped onto the Ising model, in which the cost function is recast as an energy landscape. This landscape is described by the Ising Hamiltonian
\begin{equation}
H = -\frac{1}{2}\sum_{i,j} J_{ij} \sigma_i \sigma_j + \sum_i h_i \sigma_i,
\label{eq:H_energy}
\end{equation}

\noindent where $\sigma_i \in \{\pm1\}$ are the binary spin variables,  $J_{ij}$ the coupling between spin $i$ and $j$ and $h_i$ a local bias field. Solving the optimization problem is then equivalent to finding the spin configuration that minimizes $H$. The IM realizes this by physically instantiating a network of coupled binary spins whose dynamics are designed to relax toward low-energy configurations of the Ising landscape. In our implementation, each spin is encoded as an analog optical intensity $x_i$ that can be converted to a binary spin through $\sigma_i = \text{sign}(x_i)$, the coupling weights $J_{ij}$ are implemented digitally, and a gradient-descent-based feedback signal drives the spin dynamics toward the ground state (see Methods section A for more information), though convergence to the true ground state is not guaranteed for any practical IM. The resulting steady-state spin configuration therefore provides an approximate solution to the original problem, or the exact solution in the best case.

The setup, illustrated in Fig.~\ref{fig:SLM_setup}(a), comprises four core components: a light-emitting diode (LED), a polarizer (POL), a spatial light modulator (SLM), and a high-resolution CMOS camera, together with lenses that collimate the LED emission and image the SLM plane onto the camera. Unpolarized light from the LED becomes linearly polarized after the polarizer and is then reflected off the SLM, which imposes a spatially varying phase shift on one polarization component of the field. This phase modulation modifies the polarization state of the reflected beam. When passing back through the same polarizer, the change in polarization is converted into an intensity modulation. The resulting intensity pattern is directed via a beamsplitter onto the CMOS camera, where it is recorded and used to compute the gradient-descent-based feedback for the next iteration. For more information on the operational principle see Methods section B and Supplementary information 1.

Figure~\ref{fig:SLM_setup}(b) illustrates the three steps of this feedback loop in more detail. First, the CMOS image reads out the intensity of all the spins in a single frame. Secondly, the spin intensities are processed digitally to compute the feedback signal $f_i$ for each spin $i$:
\begin{equation}
f_i = \alpha x_i + \beta \sum_j J_{ij} x_j
\label{eq:feedback_setup}
\end{equation}

\noindent where the feedback strength $\alpha$ and the coupling-strength $\beta$ are tunable hyperparameters. Third and finally, the resulting feedback signal is then encoded as a phase pattern and written to the SLM, which together with the polarizer, updates the optical intensity field that is measured on the CMOS camera, thereby closing the feedback loop.

Because every spin occupies a physically distinct region of the SLM, the total number of spins is set directly by the SLM's pixel count rather than by a time-division budget, in contrast to time-multiplexed IMs \cite{bohm2019poor, 200GOPS_IM_2025, Zhenhua2024OIM, Cen2022CIM}. Concretely, each spin can be encoded on a single SLM pixel, with a surrounding pixel left unused to prevent optical crosstalk between neighboring elements, see also Supplementary information 3. Our IM can therefore in principle be extended to millions of spins simply by increasing SLM and camera resolution, without any change to the feedback architecture, giving it a direct path to scale. Furthermore, because the signal modulation relies on polarization-encoded changes that map to intensity rather than on phase shifts, as in MZMs, the system exhibits high robustness against temperature changes and maintains stable operation over measurement runs exceeding 12\,h (see also Supplementary information 4).


\begin{figure*}[t!]
    \centering
    \includegraphics[width=\textwidth]{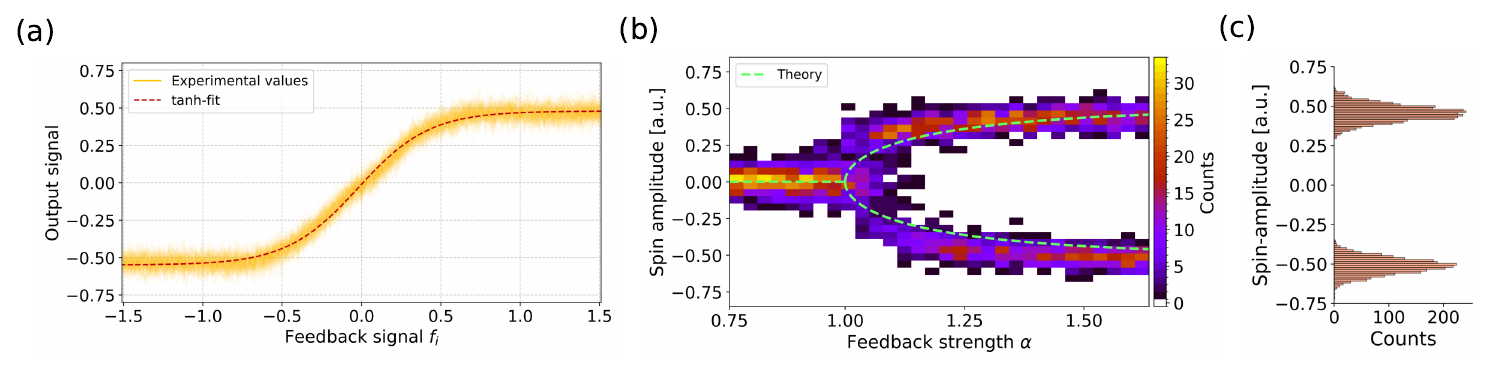}
    \caption{\textbf{Characterization of the ASIM.} 
    \textbf{(a)} The yellow curves show the experimentally measured response functions for 100 spins, plotted on top of one another. The red dashed line represents a hyperbolic tangent function, which closely matches the experimental data. This agreement demonstrates that the clipping dynamics in the hardware are well described by this nonlinearity. 
    \textbf{(b)} Experimentally measured distribution of 100 uncoupled spins, shown together with the theoretically expected pitchfork bifurcation in green. The applied feedback strength on the x-axis is normalized such that the bifurcation occurs at $\alpha = 1$.
    \textbf{(c)} Distribution of the final spin amplitudes for a 100-spin system after 100 iterations with $\alpha = 1.61$ and for 50 runs.
    }
    \label{fig:IM_characterization_v1}
\end{figure*}

\subsection{Ising machine characterization}

Different nonlinear response functions can be used in IMs \cite{bohm2021order}. Due to clipping, see Methods B for more information, our system is best described using a hyperbolic-tangent-like response function which we model in discrete time as
\begin{equation}
    \label{eq:respondse_function}
    x_i[k+1] = \tfrac{1}{2} \tanh\big(f_i[k] + \zeta_i[k]\big)
\end{equation}
where $k$ denotes the iteration number, $f_i$ is the feedback signal defined in Eq.~\ref{eq:feedback_setup}, and $\zeta_i$ represents Gaussian white noise.

This behaviour is also confirmed experimentally in Fig.~\ref{fig:IM_characterization_v1}(a), which shows in yellow the measured response function for 100 independent spins as a function of the applied feedback signal $f_i$. The resulting response function closely follows a $\tanh$ response, as highlighted by the fitted curve, indicated by the dotted line, yielding a high $R^2 = 0.9988$ and a low RMSE $=0.0150$. Further information is provided in Methods section C. This specific nonlinearity has been shown to offer favorable spin homogeneity, straightforward bias implementation, and robust noise tolerance relative to alternatives \cite{bohm2021order, de2025incorporating, mys2026predicting}.

Next, we characterize the bifurcation behaviour of the spin variables. A pitchfork bifurcation underlies a large class of IM implementations, as it enables uncoupled spins ($\beta = 0$) to evolve from a single zero-amplitude fixed point into two states, corresponding to spin-up and spin-down, once the feedback strength $\alpha$ exceeds a threshold. This transition provides the basis for mapping continuous spin amplitudes onto discrete binary states and is therefore a key operating principle underlying our analog IM implementation.

We assess this behaviour in the ASIM by measuring the bifurcation directly on our hardware. Figure~\ref{fig:IM_characterization_v1}(b) shows the amplitude distribution of 100 uncoupled spin variables after 100 iterations at different feedback strength values (normalized so that the bifurcation point falls at $\alpha = 1$). For $\alpha$-values below the bifurcation point, all spins relax to zero amplitude, while above it, this trivial solution becomes unstable and the spins split into two stable, non-zero branches that follow the expected theoretical pitchfork form, shown in green. (Further information is provided in Methods section D). Fig.~\ref{fig:IM_characterization_v1}(c) shows the resulting bimodal amplitude distribution at $\alpha = 1.61$ across 50 independent runs, confirming that the two bifurcation states are equally probable and well separated.

\subsection{Benchmark results}

Having characterized the uncoupled dynamics, we next benchmark the machine's performance as a solver. Its reconfigurability lets us test it across problem classes without any hardware changes, using two standard combinatorial benchmark types. The first is the maximum-cut (Max-Cut) problem, in which the vertices of a weighted graph are partitioned into two sets according to a spin assignment $\{\sigma_i\}$, and the IM seeks to maximize the cut value, which is the total weight of edges $(i,j)$ whose endpoints fall into opposite sets. The second is the number partitioning problem (NPP), in which a set of integers $\{a_i\}$ is split into two subsets according to $\{\sigma_i\}$, and the IM seeks to minimize the absolute difference between the two subsets' sums, which is referred to as the partition value. Both problems are NP-hard in general and map directly onto the Ising formulation \cite{200GOPS_IM_2025, sevenants2025requirements}. A detailed discussion of this mapping is provided in Methods section  E \& F. The hyperparameters ($\alpha$, $\beta$) are optimized separately for each benchmark problem using brute-force parameter scans in simulation (see Supplementary information 5 for details). In hardware, we additionally implement a linear annealing schedule for $\beta$ for spare problems while for dense graphs we use a mixing term $h$, introduced in Ref.~\cite{sevenants2026reducing}. This mixing term reduces the IM's sensitivity to the choice of hyperparameters, which is especially useful for more densely connected problems, making it easier to identify parameter settings that consistently drive the IM to low energy states (see Supplementary information 5 for details).

\begin{figure*}[t]
    \centering
    \includegraphics[width=\textwidth]{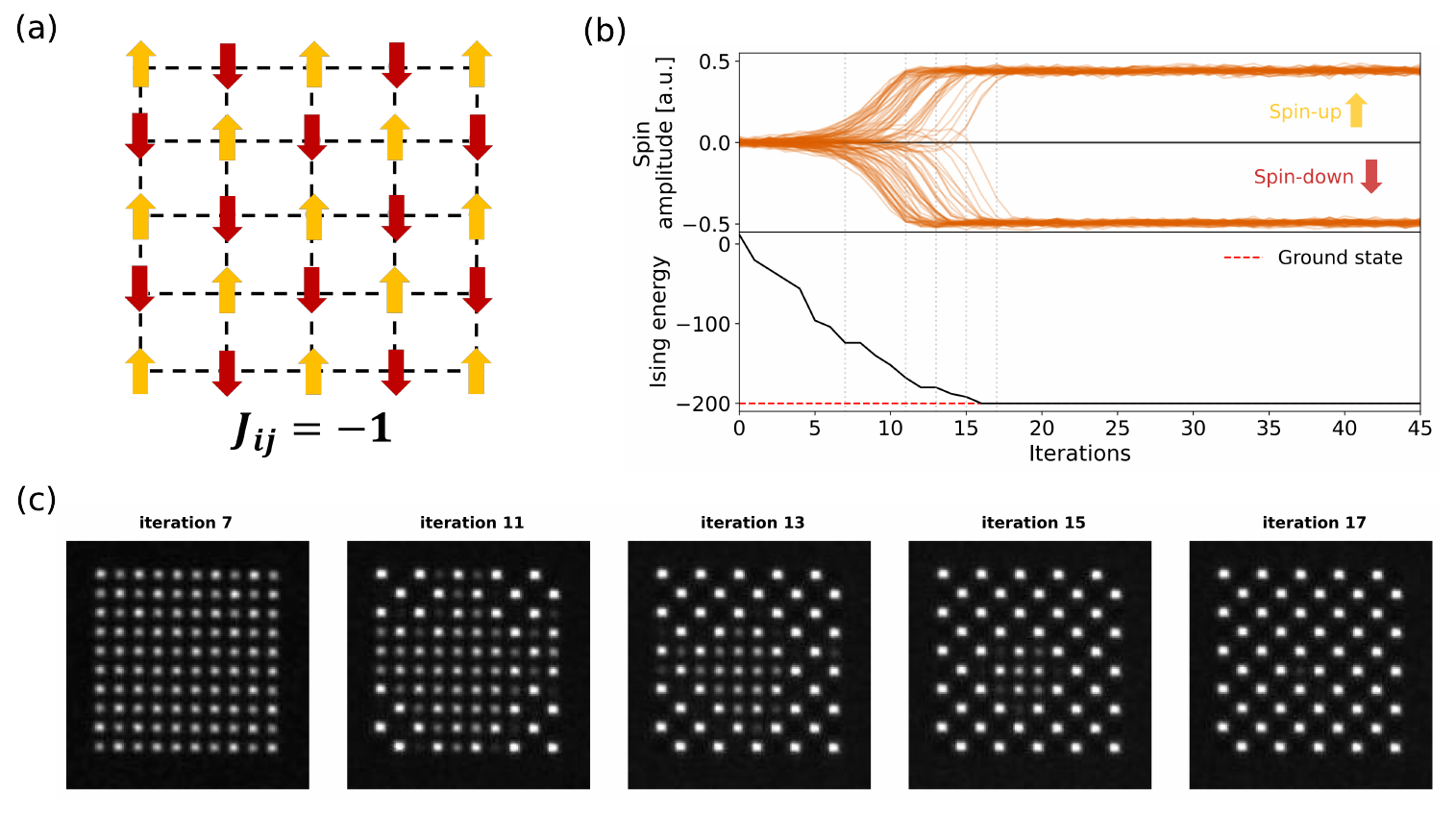}
    \caption{\textbf{Time-evolution of the ASIM on a square lattice with nearest-neighbor, anti-ferromagnetic coupling.} 
    \textbf{(a)} Illustration of the ground-state configuration. The red color-spins are spin-down while the yellow spins indicate the spin-up. 
    \textbf{(b)} Evolution of the experimentally measured spin amplitudes and the corresponding Ising energy for a \(10 \times 10\) antiferromagnetic square-lattice problem. The ground-state solution is reached after 17 iterations, at which moment the spins settles in a stable binary state. 
    \textbf{(c)} Snapshots recorded by the CMOS camera at five different times during the evolution, as indicated by the grey vertical lines in panel~(b). Brighter squares correspond to spin-up states, while darker squares correspond to spin-down states.}
    \label{fig:SpinEvolutionExample}
\end{figure*}

We first benchmark Max-Cut performance on three problem classes: the square-lattice problem, and instances from the BiqMac and Gset libraries. The square lattice problem is a tractable special case of Max-Cut with nearest-neighbour antiferromagnetic couplings ($J_{ij}=-1$) under periodic boundary conditions, whose ground state is known a priori to be the alternating checkerboard pattern of Fig.~\ref{fig:SpinEvolutionExample}(a) and which can be found in polynomial time ~\cite{barahona1982computational}. Although this problem is not NP-hard, its tractability is what makes the square lattice a valuable benchmark. The problem size can be scaled arbitrarily while its computational hardness remains fixed, and because the ground-state energy and configuration are known exactly, solution quality can be verified directly rather than estimated. These properties make it a useful problem for assessing the maximum problem size that our setup can reliably handle.

As a first example, Fig.~\ref{fig:SpinEvolutionExample}(b) shows the measured spin-amplitude and energy evolution for a $10 \times 10$ antiferromagnetic lattice (100 spins). The spins initially fluctuate near zero before bifurcating into stable spin-up and spin-down states. The Ising energy, described through Eq.~\ref{eq:H_energy}, decreases monotonically and reaches the ground-state value within 17 iterations. Fig.~\ref{fig:SpinEvolutionExample}(c) shows the corresponding CMOS frames at selected iterations in which  the initially uniform intensity resolves into a high-contrast checkerboard pattern, fully formed by iteration 17, confirming convergence to the ground state.

The main objective of this work is to demonstrate the scaling capabilities of this novel ASIM architecture. To do this, we start from an antiferromagnetic lattice of 100 spins and scale it up by three orders of magnitude to 110{,}889 spins. Importantly, the spins are not time-multiplexed. All variables are measured simultaneously, in a single shot, with the CMOS camera.

\begin{figure*}[t]
    \centering
    \includegraphics[width=1.0\linewidth]{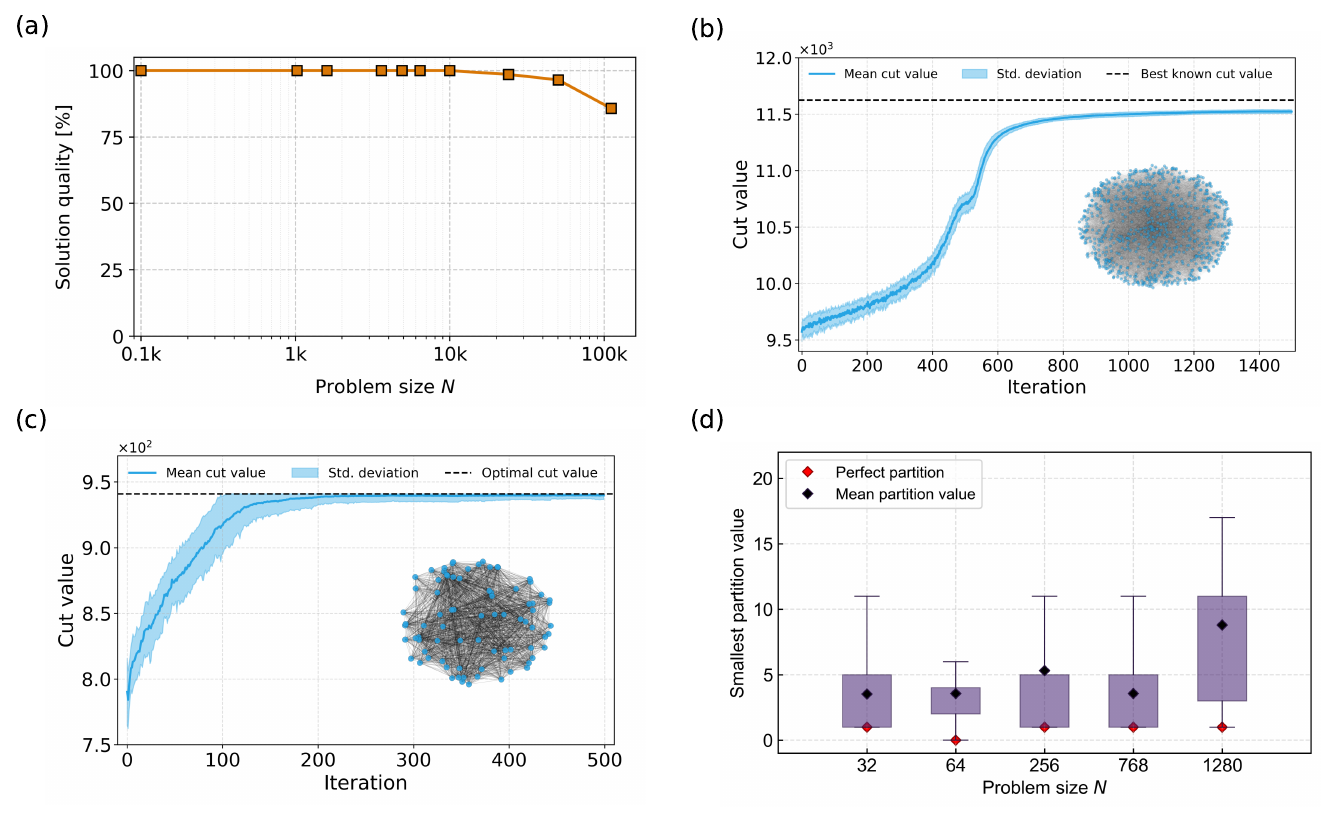}
    \caption{\textbf{Benchmark results of the ASIM.} \textbf{(a)} Solution quality as a function of the anti-ferromagnetic 2D lattice size.
    \textbf{(b)} The cut-value as a function of iteration number for the $G1$ benchmark. The dark line indicates the average value, while shaded area indicates the standard deviation. The inset visualizes the graph. 
    \textbf{(c)} The cut-value as a function of iteration number for the $g05\_80.1$ benchmark. The dark line indicates the average value, while shaded area indicates the standard deviation. The inset visualizes the graph.
    \textbf{(d)} Boxplots showing the distribution of the lowest partition values, for five different problem sizes of the number partitioning problem. Dark markers indicate the mean value for each size. The red dots indicate the perfection partition for each problem size. 
    }
    \label{fig:benchmark_results}
\end{figure*}

Figure~\ref{fig:benchmark_results}(a) shows the solution quality, defined as the fraction of the (best-known) ground-state energy recovered by the IM, as a function of problem size on a logarithmic $x$-axis. Solution quality is averaged over 50 independent runs per problem size, except for the largest instance of 110,889 spins, where memory and runtime constraints limited us to 20 runs. The solution quality remains at $100\%$ up to 10,000 spin variables, after which it drops slightly to around $86\%$ for the largest problem size considered. This decrease is consistent with literature and is not caused by a hardware limitation, but rather by slower domain-wall removal in larger lattices \cite{200GOPS_IM_2025, bohm2019poor}. Domain walls separate regions that are locally antiferromagnetically ordered but misaligned with each other, and larger systems typically require more iterations to remove them completely.

In addition to evaluating the scaling behavior of the system, we also benchmark its performance on more challenging Max-Cut problems drawn from the publicly available Gset and BiqMac libraries \cite{BiqMacLibrary, GsetLibrary}. We have taken one from each of them as the former has sparse graphs while the latter has more dense graphs. 

We start with the sparse G1 benchmark from the Gset library, for which Fig.~\ref{fig:benchmark_results}(b) shows the average cut value over 50 runs as a function of iteration number. The shaded area indicates the standard deviation, while the inset illustrates this 800-spin graph. Since the exact optimum is generally unknown for Gset instances, performance is evaluated relative to the best-known cut value reported in the literature, shown as the horizontal dotted line. Although this value is not reached in our experiments, the IM achieves 99.43\% of the best-known cut value, comparable to what has been reported in the literature for the G1 benchmark on opto-electronic analog IMs \cite{liao2024overdamped, li2021accuracy}.

Next, we consider the denser $g05\_80.1$ BiqMac benchmark, which has an edge probability of $0.5$. Figure~\ref{fig:benchmark_results}(c) shows the average cut value over 50 runs as a function of iteration number, with the shaded area again indicating the standard deviation and the inset again showing the corresponding graph. Since this is a dense problem, we use a mixing term $h=0.5$, The cut value reaches the optimum after 100 iterations, while the decreasing standard deviation indicates that the IM consistently converges to low-energy states. After 500 iterations, we found that the probability of reaching the ground-state is 84\%, which is consistent with prior work \cite{sevenants2026reducing}.

Together, the BiqMac and Gset results show that our IM finds optimal or near-optimal solutions across both dense and sparse Max-Cut instances, with a tight run-to-run spread indicating reliable rather than occasional convergence.

To test generality beyond Max-Cut, we further benchmark our IM on five randomly-generated NPPs, which involve weighted couplings \cite{mertens2006number, hayes2002computing}, for problem sizes $N \in [32, 64, 256, 768, 1280]$ (see Methods section F for more details). As mentioned before, the objective is to divide a set of numbers into two subsets such that the difference between their sums is minimized. If this difference is $\leq 1$, one refers to it as a perfect partition. Besides being a well-studied academic benchmark, the NPP is also of practical relevance, with applications in areas such as resource allocation in manufacturing, job scheduling in multi-core processors, and cryptography, where related partitioning concepts have been used in the design of secure public-key encryption systems \cite{coffman1991probabilistic, tsai1992asymptotic, merkle1978hiding}. Benchmarking it alongside Max-Cut demonstrates that our setup is not tied to a single problem structure. For transparency and reproducibility, we have made these NPP instances that are used in this work publicly available (see Data availability).

Figure~\ref{fig:benchmark_results}(d) shows our results for this benchmark, showing the distribution of the partition value across 50 independent runs as a function of problem size. The black diamonds indicating the mean of each distribution and outliers are omitted for clarity. Owing to the very dense connectivity of this problem class, we used a mixing term of $h = 0.1$ to increase the probability of reaching a small partition value.
As seen in Fig.~\ref{fig:benchmark_results}(d), we found the perfect partition for each problem. The average value and the spread of the distribution both begin to increase for the largest problem sizes. We attribute this to the property of the NPP that the energy gap between the optimal and near-optimal partitions shrinks exponentially as the number of elements grows, hence requiring correspondingly finer numerical precision to resolve the true optimum \cite{mertens2006number}. Since we use a fixed signal-to-noise ratio (set by the SLM and CMOS camera) for all the problem sizes, an increasing share of individual runs converge to a near-optimal rather than exact solution as the required precision exceeds this margin, widening the distribution. 

\subsection{Run parallelization}

\begin{figure*}[t!]
    \centering
    \includegraphics[width=\linewidth]{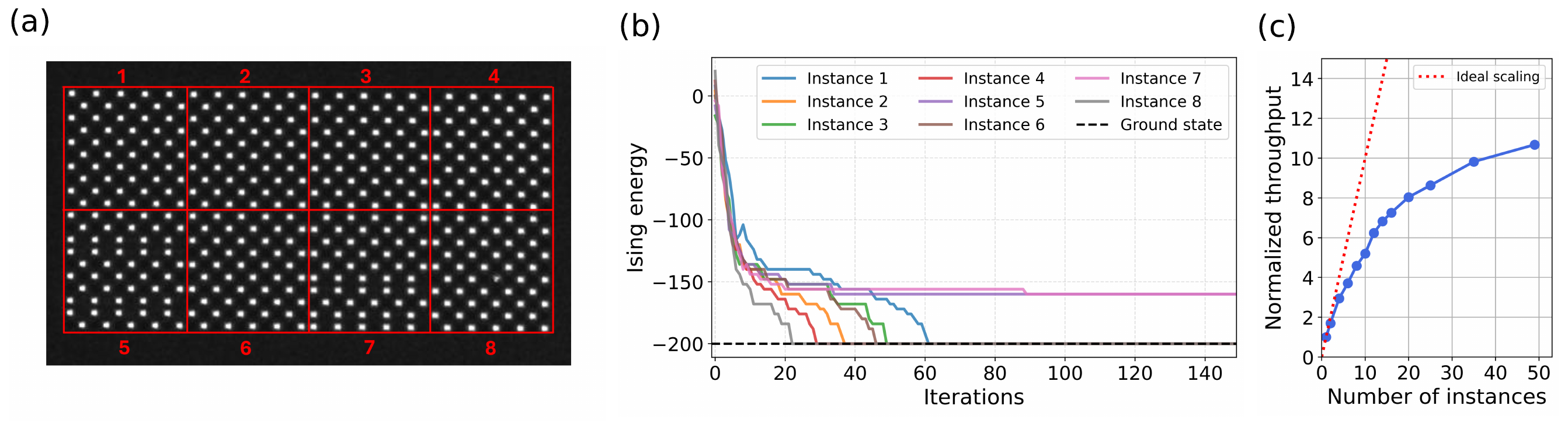}
    \caption{\textbf{Parallel runs on the ASIM.} \textbf{(a)} CMOS image showing eight independent instances running in parallel, each solving the 10 $\times$ 10 square lattice problem. Red lines and numbers are added for visual clarity.
    \textbf{(b)} Energy evolution of the eight instances, demonstrating the independent dynamics for each run. 
    \textbf{(c)} The normalized throughput as function of the number of problem instances programmed on the SLM. The problem used here is the 10 $\times$ 10 square lattice problem problem.
    }
    \label{fig:MultipleRuns}
\end{figure*}

For smaller problem sizes, a significant fraction of the SLM area is unused. Because spins are addressed spatially rather than sequentially, this idle area can be filled with additional, independent problem instances rather than left unused‚ an option that has no equivalent in time-multiplexed architectures. To demonstrate this, we implement again the $10 \times 10$ square lattice problem and replicate it eight times across the SLM, with each instance evolving independently. A CMOS image of this configuration is shown in Fig.~\ref{fig:MultipleRuns}(a), where red lines and labels are added for the reader to visually differentiate the separate lattices. Each instance receives an independent feedback signal, resulting in fully decoupled dynamics, as illustrated by the distinct energy evolutions in Fig.~\ref{fig:MultipleRuns}(b).

To quantify the performance improvement obtained by implementing multiple problem instances on the SLM in a manner directly relevant for IM, we define the throughput TP as the total number of spin updates performed per second:
\begin{equation}
\label{eq:througput}
TP = \frac{N \cdot I}{T_{R}},
\end{equation}
where $N$ is the total number of spins displayed on the SLM, $I$ is the number of iterations in a single run, and $T_{R}$ is the time required to complete a single run. This metric is thus problem-independent and captures the underlying architectural advantage of the platform, regardless of the particular solution strategy employed. For this reason, it is important that $T_R$ captures the complete end-to-end runtime of the feedback loop, rather than only the SLM's modulation rate, ensuring that $TP$ reflects any bottleneck present in the full pipeline rather than one that would remain hidden if only the SLM modulation rate were considered.

Fig.~\ref{fig:MultipleRuns}(c) shows the normalized throughput, defined relative to the throughput of a single instance. The dashed line represents ideal linear scaling, corresponding to the case where no overhead is introduced. Doubling the number of parallel instances then doubles the number of spin updates obtained per round trip without doubling the time required, and therefore doubles the throughput. From the figure, it can be seen that the number of spin updates per second can be increased by an order of magnitude by running 35 parallel problem instances on the SLM. However, near-ideal scaling is only achieved for a limited number of instances, after which the normalized throughput begins to saturate. Since all feedback signals were computed here on a single CPU core, this saturation reflects the serial computation bottleneck. 


In Supplementary information 6 we show a more detailed analysis which reveals that this deviation from ideal scaling originates from the increased computational cost of the digital feedback processing. Since the feedback signal for each parallel instance must be calculated independently, this processing step is inherently suitable for parallelization. To better utilize our available computational resources, we also show in Supplementary information 6 that we can distribute the feedback computation over multiple CPU cores. Specifically, we assign one CPU core per problem instance programmed on the SLM. With this parallelized feedback implementation, an order-of-magnitude increase in throughput is achieved already with 14 parallel instances, compared to the 35 instances required in Fig.~\ref{fig:MultipleRuns}(c) when using a single CPU core for feedback processing.

The independence of the parallel instances makes this approach particularly well suited for smaller problem sizes, where multiple instances can be placed side by side. This provides a key advantage over time-multiplexed architectures, as multiple problems can be processed simultaneously rather than sequentially. Furthermore, the architecture allows different types of problems to be solved in parallel. Finally, this parallel layout is also naturally compatible with parallel tempering schemes.


\section{Discussion}

\begin{figure*}[t!]
    \centering
    \includegraphics[width=1.0\linewidth]{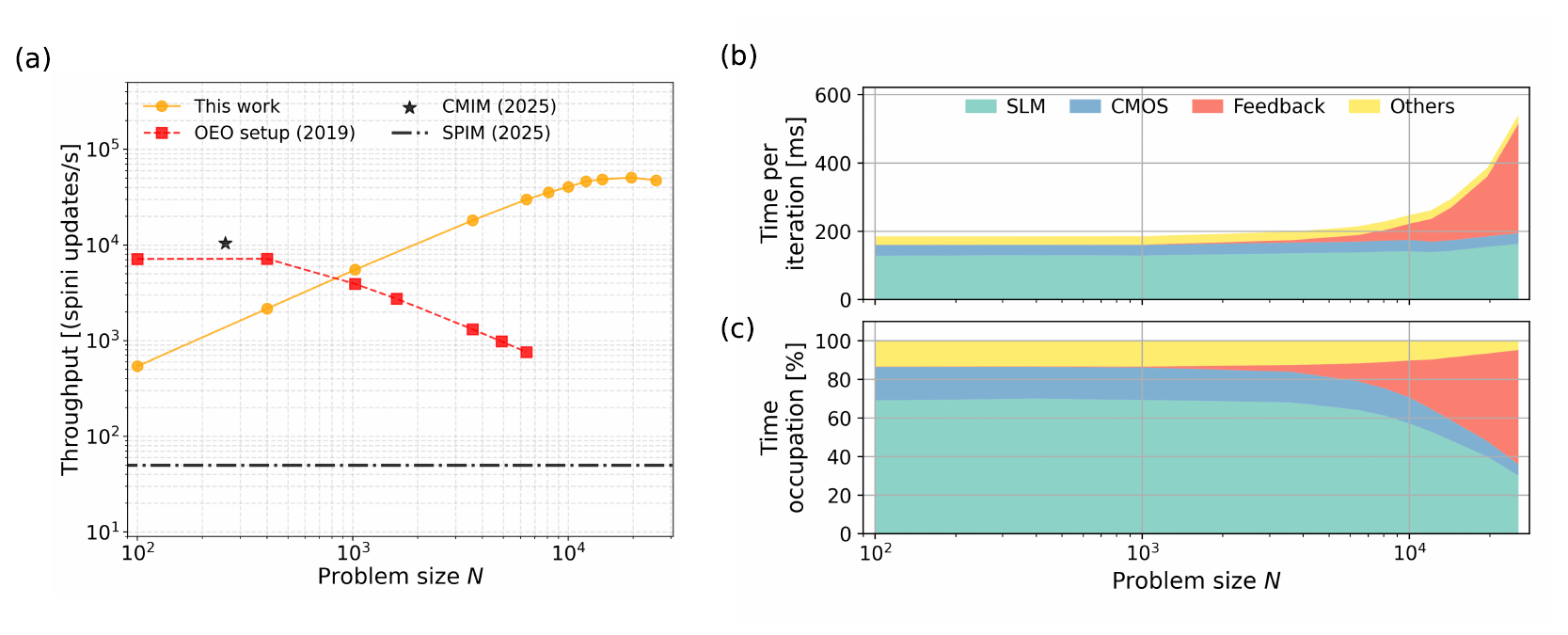}
    \caption{\textbf{Throughput and time budget of the ASIM as a function of problem size.}
    \textbf{(a)} Scaling of the throughput, defined as the number of spin updates per second, with the problem size for fully connected problems. Red squares show values measured on the time-multiplexed OEO-IM \cite{bohm2019poor}, while the black star is taken from ref.~\cite{200GOPS_IM_2025}. The black dotted line shows the throughput reported for the SPIM \cite{veraldi2025SPIM}. Yellow circles show the throughput achieved in this work.
    \textbf{(b)} The measured total time per iteration, in milliseconds, decomposed into the four main stages, addressing the SLM, reading out the CMOS, computing the feedback, together with the remaining overhead (others). The SLM and CMOS contributions vary only weakly with problem size, so that the SLM dominates at small problem sizes, whereas the feedback calculation becomes the main bottleneck at large problem sizes.
    \textbf{(c)} The same decomposition, expressed as the relative contribution to the total time per iteration, in percent.}
    \label{fig:time_periteration_scaling}
\end{figure*}

In this work, we present an analog spatially-multiplexed IM capable of handling up to 110{,}889 spins which, to the best of our knowledge, is the largest analog IM demonstrated to date. Since scalability is our primary focus, a key figure of merit in this context is how the computational cost scales with system size. Figure~\ref{fig:time_periteration_scaling}(a) shows the throughput, as defined in Eq.~\ref{eq:througput}, of our system compared against different IMs as a function of problem size, including the throughput reported for the time-multiplexed OEO-IM \cite{bohm2019poor}, the more recent, state-of-the-art cascaded-modulator Ising machine (CMIM) \cite{200GOPS_IM_2025} and the spatial photonic Ising machine (SPIM) \cite{veraldi2025SPIM}. All values are computed for a fully connected coupling matrix, meaning the full matrix-vector multiplication (MVM) is performed in the feedback loop without any sparsification or parallelization. Note that the OEO-IM relies on FPGA-based processing, and the CMIM on state-of-the-art photonic components, while our current implementation runs on a CPU in Python.

From Fig.~\ref{fig:time_periteration_scaling}(a) it can be seen that our system exhibits favorable scaling, meaning that the throughput increases with problem size, whereas time-multiplexed architectures such as the OEO-IM show the opposite trend, resulting in a two order-of-magnitude difference at the largest sizes considered. Although the OEO-IM is an older system, we have direct access to it, enabling extensive benchmarking under identical conditions up to the point where memory limitations on that system prevented testing larger problem sizes. Moreover, its time-multiplexed architecture, the key feature relevant to this comparison, is shared by many current implementations, including the CMIM. For this latter system, we only have 1 data point from literature \cite{200GOPS_IM_2025}. Figure~\ref{fig:time_periteration_scaling}(a) also includes the throughput for the SPIM, whose hardware is closely related to ours \cite{veraldi2025SPIM}. It likewise uses an SLM and a CMOS camera, but unlike in this work, the CMOS readout there is used to evaluate the Ising Hamiltonian, while a digital heuristic, e.g. Metropolis--Hastings-style simulated annealing, proposes a binary spin flip, which is accepted with a probability that depends on the change of energy caused by the spin flip. Importantly, this scheme flips only a single binary spin (or a small, fixed number of binary spins) per iteration, independent of problem size, since flipping more binary spins simultaneously typically reduces the acceptance probability and stability of the heuristic. As a result, the SPIM's throughput does not increase with problem size but instead remains constant. 

To understand the scaling of our system in Fig.~\ref{fig:time_periteration_scaling}(a), we analyze the time budget of our system. Figures~\ref{fig:time_periteration_scaling}(b) and (c) show the absolute and relative time distribution, respectively, for our spatially multiplexed IM. This total time consists of the time for addressing the SLM, reading out the CMOS, and computing the feedback, together with the remaining overhead that occurs throughout the feedback loop. The SLM and CMOS contributions vary only weakly with problem size since in our system all spin amplitudes are acquired and updated in parallel in a single shot. Therefore, the total read-out and update time stays approximately constant regardless of system size, so a larger problem spreads this fixed cost over more spins, and throughput in Fig.~\ref{fig:time_periteration_scaling}(a) rises with problem size. Time-multiplexed approaches, in contrast, address spins sequentially, so a fixed read-out time per spin causes their throughput to fall as the problem size grows.

This favorable scaling is not unlimited, however. At the largest problem sizes shown here, all for a fully connected coupling matrix, Fig.~\ref{fig:time_periteration_scaling}(b) shows that the total time per iteration itself starts to increase, which flattens the throughput curve in Fig.~\ref{fig:time_periteration_scaling}(a) rather than letting it continue to rise. This increase in time per iteration is due to the time needed to calculate the feedback signal, which will ultimately limit the TP. We show in Supplementary information~7 that for sparse coupling matrices, sparsifying the feedback computation removes this bottleneck almost entirely and restores favorable throughput scaling beyond the 25{,}600-spin limit reached here, which is set by memory constraints when uploading a full coupling matrix.

Figure~\ref{fig:time_periteration_scaling}(b) and (c) reveal two distinct regimes. For small problems ($N < 10^4$), the time per iteration is essentially invariant with problem size. For larger problems ($N \geq 10^4$), the time per iteration begins to scale with $N$, driven by the growing contribution of the feedback calculation, specifically the MVM (see also Supplementary information~7 for more details). We discuss these two regimes separately below.

For small problems ($N < 10^4$), Fig.~\ref{fig:time_periteration_scaling}(b) and (c) show that addressing the SLM is the dominant contribution to the time budget. This bottleneck is shared with the SPIM architecture \cite{pierangeli2019SPIM, fang2021SPIM, ye2023SPIM, veraldi2025SPIM}. Since the throughput of the SPIM’s does not increase with problem size, it remains bound by SLM addressing at every problem size we consider. Our spatially-multiplexed setup however can substantially reduce this SLM bottleneck at small problem sizes by programming multiple independent SLM instances in parallel on a single SLM, a strategy already demonstrated in Fig.~\ref{fig:MultipleRuns}. Running multiple optimizations concurrently in this way yields more than an order-of-magnitude increase in throughput, largely closing the gap to time-multiplexed systems visible for the smallest problem sizes in Fig.~\ref{fig:time_periteration_scaling}(a).

For larger problems ($N \geq 10^4$), we see from Fig.~\ref{fig:time_periteration_scaling}(b) and (c) that not the SLM but the feedback calculation is the dominant contribution to the time budget. Unlike the  SLM bottleneck at small problem sizes, this feedback cost cannot be mitigated by running independent problems in parallel, since it scales with the size of a single problem instance. Our free-space optical architecture nevertheless offers a natural path forward. Since the growing time cost is caused by the MVM required for the feedback calculation, this computation could instead be performed by high-end digital hardware in the future. An even more promising path is to offload the MVM to the optical domain rather than performing digitally \cite{spall2020fully, latifpour2024hyperspectral, chen2025free, goodman1978fully, zhou2022photonic}, removing the bottleneck altogether and allowing throughput to keep scaling favorably with problem size rather than saturating as in Fig.~\ref{fig:time_periteration_scaling}(a). In particular, an incoherent implementation of this optical MVM would avoid the interferometric phase stability required by coherent approaches, and is a natural extension of the intensity-based encoding and readout already used for the spin variables in this work \cite{goodman1978fully, zhou2022photonic}.

Our setup thus demonstrates that large-scale analog spin systems can be implemented and operated reliably in a simple and robust free-space architecture, while achieving performance competitive with state-of-the-art systems. 


At present, our system is limited to approximately 110{,}889 spin variables, set by the field of view of the imaging system, and draws approximately 100~W of power (see Supplementary information~8 for the full breakdown). Beyond this, the achievable scale is set primarily by the pixel count of the SLM and the resolution of the CMOS sensor. As both technologies continue to improve, we expect scaling to systems with more than $10^6$ spins to be feasible with suitable hardware.

\section{Conclusion}
We have demonstrated a novel spatially multiplexed analog IM capable of handling up to 110{,}889 analog spin variables. The free-space architecture, built from off-the-shelf components, enables robust and scalable operation while naturally supporting the parallel readout and update of all spins, achieving performance comparable to the current state of the art. We further show that the spatially parallel nature of this novel setup gives rise to a favorable scaling in which the throughput increases with system size, in contrast to the more common time-multiplexed approaches. At the same time, it also allows for multiple problem instances to be executed simultaneously, increasing computational throughput by at least an order of magnitude, another property that is non-existing for time-multiplexed setups. 

At present, the achievable problem size is primarily set by the field of view of the imaging system, which limits the number of spins that can be simultaneously captured by the camera. This is a practical rather than a fundamental constraint as a refined imaging setup could relax this in the future, potentially scaling the system toward millions of spins. Combined with further improvements in dedicated hardware components and the integration of optical matrix--vector multiplication, the performance and scale of the system can be enhanced significantly further. More broadly, our results highlight the potential of spatial photonic computing as a scalable and accessible platform for large-scale optimization, marking a step towards practical, large-scale photonic IMs.

\section{Data availability}
The NPP problems used in this work are available through the GitHub page: \url{https://github.com/VUB-APHY/Number-Partitioning-Problems}. The authors declare that all other relevant data are included in the manuscript. Additional data are available from the corresponding authors upon request.

\section{Author contributions}
T.S. built the setup, wrote the main code, performed the benchmark experiments and the throughput-scaling analysis, and wrote the manuscript. J.V.C. helped write the main code. K.W. wrote the code for parallel runs and helped benchmarking them. P.B., G.V.d.S. and G.V. supervised the project and also provided feedback on the manuscript. All authors discussed the results and reviewed the manuscript.

\section{Additional information} 
The authors would like to thank Stijn Van Vooren, Jacob Lamers, Leen Mys and Yunfeng Nie for the interesting discussions and feedback. The authors additionally wish to acknowledge Ryo Mizuta Graphics for supplying the freely available set of 3D optical components that were employed in Fig.\ref{fig:SLM_setup}. This research was funded by the Research Foundation Flanders (FWO) under grant G0A6L25N. Additional funding was provided by the EOS project ”Photonic Ising Machines”. This project (EOS number 40007536) has received funding from the FWO and F.R.S.-FNRS under the Excellence of Science (EOS) programme.

\printbibliography[
  heading=bibliography,
  title={References}
]


\section{Methods}

\subsection{Energy landscape navigation}
To navigate the Ising energy landscape, we replace the discrete spins $\sigma_i \in \{\pm1 \}$ by continuous analog variables $x_i$, with $\sigma_i = \operatorname{sign}(x_i)$. The analog energy landscape is defined by
\begin{equation}
\label{eq:analog_hamiltonian}
H_{analog} = -\frac{1}{2}\sum_{i,j} J_{ij} x_i x_j + \sum_i h_i x_i.
\end{equation}
At each iteration, the feedback applied to the system is obtained from a weighted combination of the current state and the gradient of the Hamiltonian,
\begin{align}
\label{eq:feedback_signal}
f_i &= \alpha x_i - \beta \frac{\partial H_{analog}}{\partial x_i} \nonumber \\
    &= \alpha x_i + \beta \sum_j J_{ij} x_j - \beta h_i,
\end{align}
where the feedback strength $\alpha$ and the coupling strength $\beta$ are tunable hyperparameters. This feedback is then applied to the SLM for the next iteration, resulting in the $x_i[k+1] = \tanh\left(f_i[k] + \gamma \zeta[k] \right)$. The saturating nonlinearity of $\tanh$ bounds the spin amplitudes and, together with the feedback strength $\alpha$, gives rise to the bifurcation behaviour discussed in the main text, while the feedback rule as a whole allows for a continuous relaxation of the discrete Ising problem, steering the dynamics toward low-energy configurations. Noise originating from the hardware, as well as noise introduced digitally, provides a stochastic perturbation that helps the system escape local minima.

\subsection{Optical hardware and calibration procedure}

\begin{figure*} [t!]
    \centering
    \includegraphics[width=\linewidth]{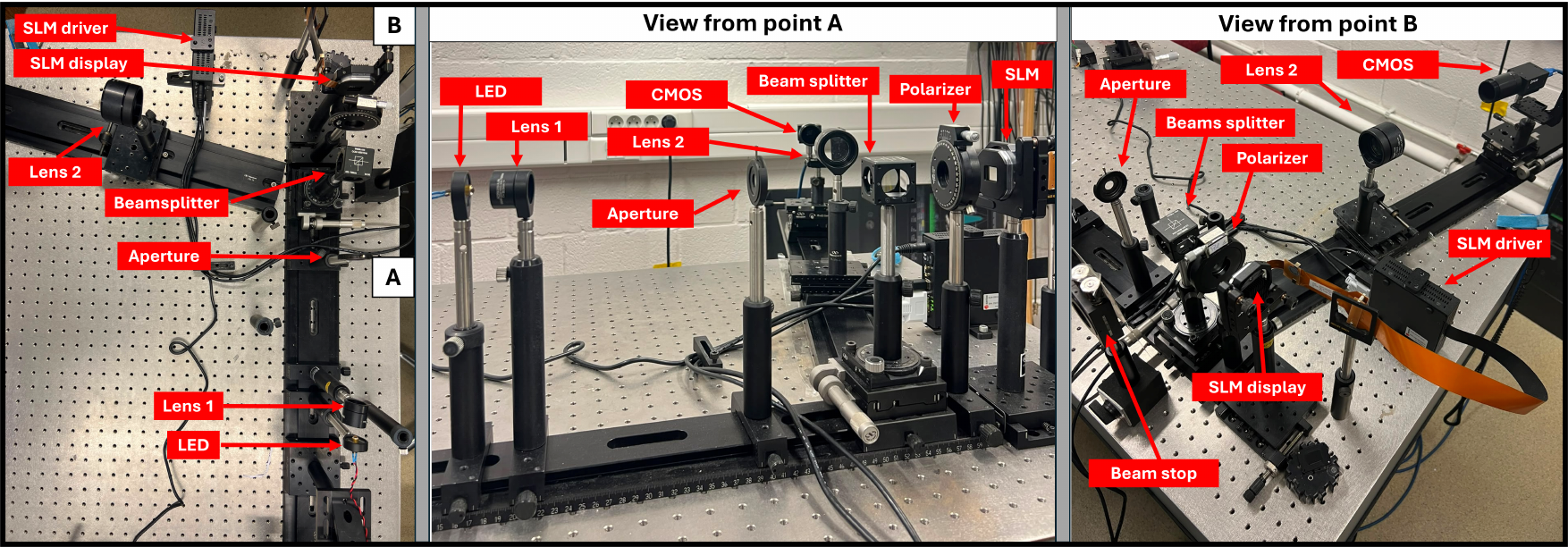}
    \caption{\textbf{Detailed layout of the experimental setup.} The left panel shows a top view of the tabletop setup, with all components labelled and indicated by arrows. The CMOS camera is omitted from this view so that the remaining components can be shown at sufficient magnification. The middle and right panels show the setup as seen from viewpoints A and B, respectively, which are indicated in the left panel; the CMOS camera is visible in both. The beam stop shown in the right panel is omitted from the middle panel for clarity, as it plays no role in the operation of the hardware.}
    \label{fig:ext_fig:RealSetup_1}
\end{figure*}

A more detailed outline of the setup is shown in Extended Data Fig.~\ref{fig:ext_fig:RealSetup_1}, in which the main components are the LED (Thorlabs, 5LED635L, $\lambda = 635\,\text{nm}$), a collimating lens (Thorlabs, AC254-030-A-ML), a non-polarizing beam splitter (CCM1-BS013/M), the linear polarizer, the SLM (Santec, SLM-200), a tube lens for imaging (Thorlabs, TTL180-A), and the CMOS camera (FLIR Grasshopper, GS3-U3-32S4M-C). We note that, due to the extended non-point-like emission area of the LED, the collimating lens does not produce a fully collimated beam, resulting in some residual divergence. Moreover, as derived in detail in Supplementary information 1, the amplitude modulation described in the main text actually follows a Malus-law response, given by

\begin{equation}
    x_i = \cos^2{(\Gamma/2)}
    \label{eq: theoretical_eq}
\end{equation}
\noindent where $\Gamma$ is the phase difference applied by the SLM. 
In practice, the measured response is further modified by the intensity offset and the regime of the response curve being used for the spin dynamics. Specifically, the calibration yields an intensity offset $x_\mathrm{offset}$, which is subtracted from the measured Malus-law response to center the spin amplitudes around zero. We operate the system around the inflection point on the uphill slope of the Malus-law response, such that the measured intensity is given by
\begin{equation}
    x_i = \cos^2{(\frac{\Gamma}{2} + \frac{3\pi}{2})} - x_\mathrm{offset}.
\end{equation}
The resulting response is then subjected to hard-coded clipping, which shapes the measured $\cos^2$ response into the hyperbolic-tangent-like nonlinearity used throughout the main text, where $\Gamma/2$ maps to the feedback signal $f_i$. This hard-coded clipping, together with the choice of the inflection point, is required because of the non-ideal shape of the measured $\cos^2$ response function (see Supplementary information 1 for more information).

In this setup, individual spins are defined by assigning each spin to a distinct region of the SLM, which is imaged onto a corresponding region of the CMOS sensor. Because of the resolution mismatch between the SLM and the camera, and diffraction in the imaging path that spreads intensity across spin boundaries rather than producing a sharp cutoff, the number of CMOS pixels per spin differs from the number of SLM pixels defining it. In addition, the LED's Gaussian spatial intensity profile causes the response to vary across spins, which, left uncorrected, can degrade computational performance. To address this, we developed a fully autonomous detection and calibration procedure consisting of several stages, including the position calibration, response-function measurement and normalization, and background calibration (see Supplementary information 2 for more information).

\subsection{Tangent-like response function}
To verify that the spins display a tanh-like response, we sweep the feedback signal $f_i$, as defined in Eq.~\ref{eq:feedback_setup}, from $-1.5$ to $1.5$ for each of the 100 used spins. 

To quantify how closely the spins follow the tanh-like behaviour, we computed the unweighted average and standard deviation of all measured response functions across the 100 spins, and fitted the resulting mean curve to a function of the form
\begin{equation}
    y = A{\,}\tanh(B{\,}\Gamma + C) + D,
\end{equation}
where $y$ is the output signal and $A$, $B$, $C$, and $D$ are the four fit parameters. The fit yields $A = 0.514$, $B = 2.221$, $C = 0.050$, and $D = -0.035$, with $R^2 = 0.9988$ and $\mathrm{RMSE} = 0.0150$. This fit is shown in Fig.~\ref{fig:IM_characterization_v1}(a) together with the measured signal for 100 different spins.

Since the spin output spans approximately $[-0.5, 0.5]$, 
the RMSE corresponds to only $\sim\!1.5\%$ of the full signal range. This means that the average deviation between the measured and fitted response is more than an order of magnitude smaller than the signal itself. Similarly, $R^2 = 0.9988$ means the tanh model accounts for $99.88\%$ of the variance in the response curve, leaving less than $0.12\%$ unexplained. Together, these values confirm that the tanh function provides a good description of the spin response.

\subsection{Pitchfork bifurcation}
The response function of our IM can be described using a tangent-hyperbolic function. Due to the measurement-feedback setup, the dynamics are not continuous but reduce to a time-discrete map given by Eq.~\ref{eq:respondse_function}. The fixed points of the time-discrete map cannot be found analytically, so Eq.~\ref{eq:respondse_function} has to be solved numerically using the Newton--Raphson method. This method approximates a root of a function by replacing it, at the current estimate, with its tangent line and taking the zero crossing of that tangent as the next estimate. Repeating this process often leads to rapid convergence to the root. To find the fixed points, we look for the function whose root is the residual
\begin{equation}
  f(x) \;=\; x - \tfrac{1}{2}\tanh(\alpha x),
  \label{eq:residual}
\end{equation}
which vanishes exactly when $x$ is a fixed point of Eq.~\eqref{eq:respondse_function}. The estimate is then updated as 
\begin{equation}
    x_{n+1} = x_n - f(x_n)/f'(x_n)
\end{equation}
with 
\begin{equation}
    f'(x) = 1 - \tfrac{\alpha}{2}\,\mathrm{sech}^{2}(\alpha x)
\end{equation}
and this is done independently for every value of $\alpha$.

For this method, we found convergence in fewer than ten steps for every $\alpha$ and leaves a residual $|f(x)|<10^{-15}$. The resulting branches are the solid curves overlaid on the measured histograms in Fig.~\ref{fig:IM_characterization_v1}(b). Note that the two branches do not keep growing with $\alpha$ as is the case in the standard supercritical bifurcation, but they bend over and clip just below $\pm\tfrac{1}{2}$. This follows directly from the shape of the response function of Eq.~\eqref{eq:respondse_function}. 


\subsection{Max-Cut}
The Max-Cut problem on a weighted graph can be mapped directly onto an Ising Hamiltonian by assigning each vertex a spin variable $\sigma_i \in \{\pm 1\}$. For a graph with edge weights $w_{ij}$, the cut value can then be written as
\begin{equation}
    C(\boldsymbol{\sigma}) = \frac{1}{4}\sum_{i,j} w_{ij}\left(1-\sigma_i\sigma_j\right).
\end{equation}
Expanding the right-hand side and substituting $w_{ij} = -J_{ij}$ gives
\begin{equation}
    2C(\boldsymbol{\sigma}) = \frac{1}{2}\sum_{i,j} J_{ij}\sigma_i\sigma_j - \frac{1}{2}\sum_{i,j} J_{ij}\,.
\end{equation}
Rearranging this expression and comparing it to the general form of the Ising Hamiltonian, 
we obtain
\begin{equation}
    H(\boldsymbol{\sigma}) = -2 C(\boldsymbol{\sigma}) - \frac{1}{2}\sum_{i,j} J_{ij}\,,
\end{equation}
where the second term on the right-hand side is a constant, independent of the spin configuration $\boldsymbol{\sigma}$. Therefore, maximizing $C$ is equivalent to minimizing the Ising energy. In this mapping, the coupling matrix elements are given by $J_{ij}=-w_{ij}$ for $i \neq j$, with no local fields required ($h_i=0$).

The benchmark instances used in this work are $G1$ from the Gset-library and $g05\_80.1$ from the BiqMac-library, both are publicly available online, and have binary coupling weights \cite{BiqMacLibrary, GsetLibrary}.

\subsection{Number partitioning}
In general, the number partitioning problem can be mapped onto an Ising Hamiltonian by assigning each integer $a_i$ from the set S = $\{a_1, a_2, ... a_N \}$ a spin variable $\sigma_i \in \{\pm 1\}$, indicating to which of the two subset the number belongs. The objective is to minimize the (squared) difference $\Delta$ between these subset sums, which mathematically can be written as
\begin{equation}
    \Delta = \left(\sum_i a_i \sigma_i \right)^2.
\end{equation}
A perfect partition is obtained when $\Delta(\boldsymbol{\sigma}) = 0$ if the total sum is even, and $\Delta(\boldsymbol{\sigma}) = 1$ otherwise. Expanding the above expression gives
\begin{equation}
     \Delta = \sum_{i} a_i^2 + \sum_{i \neq j} a_i a_j \sigma_i \sigma_j \, .
\end{equation}
This can be directly mapped onto an Ising Hamiltonian of the form
\begin{equation}
    H(\boldsymbol{\sigma}) = -\frac{1}{2} \sum_{i,j} J_{ij} \sigma_i \sigma_j = \Delta - \sum_{i}^{N} a_i^2\,,
\end{equation}
where the second term on the right-hand side is a constant, independent of the spin configuration $\boldsymbol{\sigma}$. Therefore, minimizing $\Delta$ is equivalent to minimizing the Ising energy. In this mapping, the coupling matrix elements are given by $J_{ij} = -2a_ia_j$  for $i \neq j$, with no local fields required ($h_i=0$). In this work, number partitioning instances are generated by randomly sampling $N$ positive integers $x$ from the range $1 \leq x \leq 2^{k}$, with $N \in [32, 64, 256, 768, 1280 ]$ and $k = 6$. The used problems are made publicly available (see Data Availability).




\newpage

\end{document}


\title{Supplementary information: spatially multiplexed opto-electronic Ising machine with single-shot parallel readout of 100,000 spins

\thanks{\hrule \vspace{0.25em}  *Correspondence and requests should be addressed to T.S. (email: toon.jan.b.sevenants@vub.be) or to G.V. (email: guy.verschaffelt@vub.be)}}

\author[*1, 2]{Toon Sevenants}
\author[1]{Joren Van Cauwenberge}
\author[2]{Kaijun Wu}
\author[2]{Peter Bienstman}
\author[1]{Guy Van der Sande}
\author[1]{Guy Verschaffelt}

\affil[1]{Applied Physics Research Group, Vrije Universiteit Brussel, Brussels, Belgium} 
\affil[2]{Photonics Research Group, Ghent University-imec, Gent, Belgium}

\maketitle

\tableofcontents
\newpage

\section{Derivation of the amplitude modulation}
Our Ising machine (IM) alters the spin amplitudes by spatially modulating the intensity of the incoming light. To do this, it uses a polarizer and a spatial light modulator (SLM). This section derives how the SLM produces this intensity modulation, and derives the resulting response function of the analog spatially-multiplexed Ising machine (ASIM) that is presented in this work. We use the Jones formalism, in which optical waves are represented as two-component vectors and linear optical elements as $2\times2$ matrices. To keep an overview, we first introduce the concept of the LC director and the SLM Jones matrix, before deriving the outgoing light intensity.

\subsection{The director}
In a liquid-crystal-based SLM (LC-SLM), the director is a unit vector describing the average local orientation of the LC molecules. By convention, the extraordinary refractive index $n_e$ is defined along the director, while the ordinary index $n_o$ applies in the plane perpendicular to it. Applying a voltage tilts the director, changing the angle $\theta$ between the director and the propagation direction, which we take here as the $z$-axis. For light propagating along $z$, the birefringent LC layer has two orthogonal linear-polarization eigenmodes. The first one is polarized in the plane spanned by the director and the propagation direction and experiences the effective refractive index $n^{(1)}(\theta)$. The second one is polarized perpendicular to this plane and experiences the ordinary index $n^{(2)} = n_o$, independent of $\theta$. These effective indices are given by
\begin{align}
\label{eq:refractive_index}
    n^{(1)}(\theta) &= \left(   \frac{\cos^2(\theta)}{n_o^2} + \frac{\sin^2(\theta)}{n_e^2}  \right)^{-1/2} \\
    n^{(2)} &= n_o \, .
\end{align}
Because $n^{(1)}(\theta) \neq n^{(2)}$ for $\theta \neq 0$, the two eigenmodes accumulate a different phase as they propagate through the LC layer.
\begin{figure}[b]
    \centering
    \includegraphics[width=1.0\linewidth]{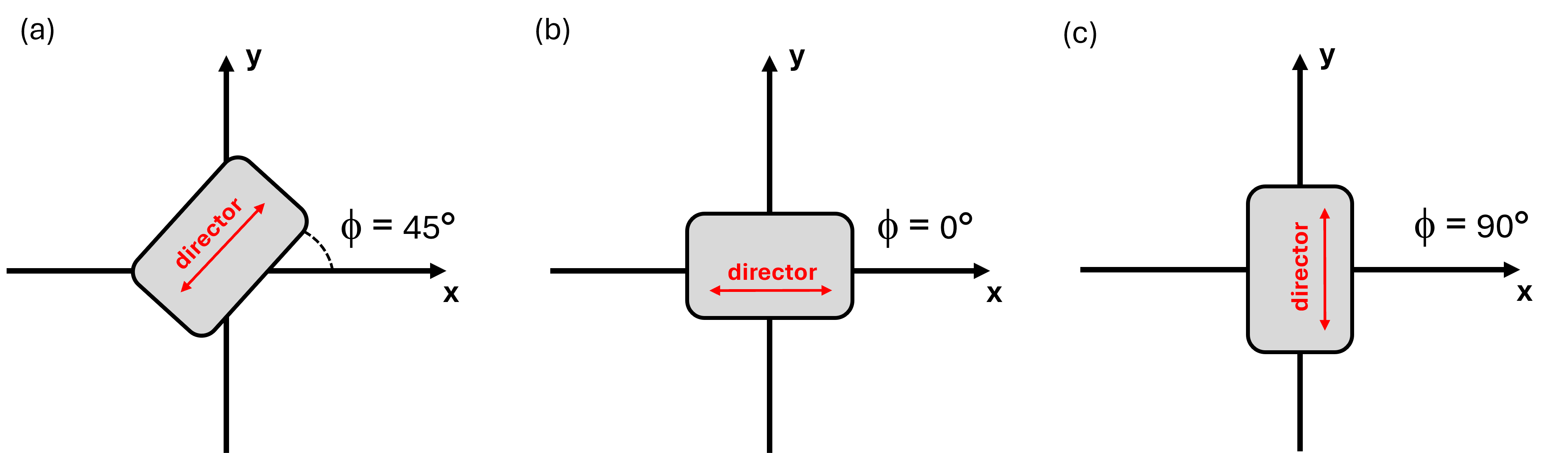}
    \caption{Different rotations of an SLM display. The projection of the LC-director in the $xy$-plane is shown in red. The azimuthal angle $\phi$ is defined relative to the $x$-axis.}
    \label{fig:azimuthal_angle}
\end{figure}

\subsection{SLM Jones matrix}
Each pixel of the SLM behaves as a birefringent element, where the exact value of the effective extraordinary index $n^{(1)}(\theta)$ is set by the local director tilt $\theta$, as given by Eq.~\ref{eq:refractive_index}. To describe the interaction of the incoming light with the SLM pixel, we also need the azimuthal angle $\phi$. Within the plane of the SLM display, the $xy$-plane, perpendicular to the propagation direction, $\phi$ is the angle between the projection of the director onto this plane and a fixed reference direction in the lab frame, the horizontal $x$-axis. Unlike $\theta$, $\phi$ is set by how the SLM is physically mounted in the optical setup and does not change with applied voltage. Different mounting orientations, and the corresponding value of $\phi$, are illustrated in Fig.~\ref{fig:azimuthal_angle}. The Jones matrix of a single SLM pixel is then given by:
\begin{equation}
    J_{SLM} = 
    \begin{bmatrix}
        \cos(\phi) & -\sin(\phi) \\
        \sin(\phi) & \cos(\phi) 
    \end{bmatrix}
    \begin{bmatrix}
        e^{-i\alpha_e} & 0 \\
        0 & e^{-i\alpha_o} 
    \end{bmatrix}
    \begin{bmatrix}
        \cos(\phi) & \sin(\phi) \\
        -\sin(\phi) & \cos(\phi) 
    \end{bmatrix}
\label{eq: J_SLM1}
\end{equation}
with 
\begin{align}
    \alpha_e &= \frac{\omega}{c} n^{(1)}(\theta) \Delta z \\
    \alpha_o &= \frac{\omega}{c} n^{(2)} \Delta z \,,
\end{align}
where $\Delta z$ is the physical path length traveled by the light through the LC layer.\newline

The first rotation matrix in Eq.~\ref{eq: J_SLM1} decomposes the incident polarization state into components along the two orthogonal (principal) axes of the SLM pixel. One along the in-plane projection of the director, which accumulates phase $\alpha_e$, and one perpendicular to it, which accumulates phase $\alpha_o$. The middle, diagonal matrix then applies this phase retardation to each component independently. Finally, the inverse rotation transforms the result back into the original laboratory frame. In our setup we have taken $\phi = 0\degree$. Note that the same measured intensity is obtained if we instead choose $\phi = 90\degree$ since this rotation swaps the roles of the two axes 
which flips the sign of the relative phase $\Gamma$ defined below. Then, since the measured intensity depends on $\Gamma$ only through $\cos^2(\Gamma/2)$ which is an even function of $\Gamma$, both choices of $\phi$ yield identical results. Both situations are illustrated in Fig.~\ref{fig:azimuthal_angle}(b) and (c) respectively. Choosing $\phi = 0\degree$ reduces Eq.~\ref{eq: J_SLM1} to 
\begin{equation}
    J_{SLM} = 
    \begin{bmatrix}
        1 & 0 \\
        0 & 1
    \end{bmatrix}
    \begin{bmatrix}
        e^{-i\alpha_e} & 0 \\
        0 & e^{-i\alpha_o} 
    \end{bmatrix}
    \begin{bmatrix}
        1 & 0 \\
        0 & 1
    \end{bmatrix}
    =
    \begin{bmatrix}
        e^{-i\alpha_e} & 0 \\
        0 & e^{-i\alpha_o} 
    \end{bmatrix} .
\label{eq: J_SLM2}
\end{equation}
\subsection{Amplitude modulation}
To achieve amplitude modulation, we place the SLM between a polarizer and an analyzer. We assume the incident light is linearly polarized at $45^\circ$ relative to the LC director, with Jones vector
\begin{equation}
    E_{in,45} = \frac{E_0}{\sqrt{2}} \begin{bmatrix} 1 \\ 1 \end{bmatrix} .
\label{eq: E_in45}
\end{equation}
First, let us focus on the light that is reflected by the SLM. This is given by:
\begin{equation}
    E_{SLM, out} = J_{SLM} E_{in, 45} = \frac{E_0}{\sqrt{2}}
    \begin{bmatrix}
        e^{-i\alpha_e} \\
        e^{-i\alpha_o} 
    \end{bmatrix} .
\label{eq: E_SLMout}
\end{equation}
Since we are only interested in the relative phase, we can rewrite the above vector by using a change of variables to
\begin{equation}
    E_{SLM, out}  = \frac{E_0}{\sqrt{2}}
    \begin{bmatrix}
        e^{-i\Gamma/2} \\
        e^{i\Gamma/2} 
    \end{bmatrix}
     e^{-i\Phi/2},
\label{eq: E_SLMout2}
\end{equation}
where $\Gamma$ and $\Phi$ are a relative and global phase respectively, and are defined as: 
\begin{align}
\label{eq:Gamma_def}
    \Gamma &= \alpha_e - \alpha_o = \frac{\omega}{c} \left( n^{(1)}(\theta) - n^{(2)} \right) \Delta z \\
    \Phi &= \alpha_e + \alpha_o = \frac{\omega}{c} \left( n^{(1)}(\theta) + n^{(2)} \right) \Delta z \,.
\end{align}
Finally, using the Jones matrix of the polarizer, we end up with:
\begin{align}
    E_{out} = J_{pol, 45} E_{SLM, out} &= \frac{E_0}{2\sqrt{2}}
    \begin{bmatrix}
        1 & 1 \\
        1 & 1
    \end{bmatrix}
    \begin{bmatrix}
        e^{-i\Gamma/2} \\
        e^{i\Gamma/2} 
    \end{bmatrix}
    e^{-i\Phi/2} \\
    &= \frac{E_0}{2\sqrt{2}} 
    \begin{bmatrix}
        e^{i\Gamma/2} + e^{-i\Gamma/2} \\
        e^{i\Gamma/2} + e^{-i\Gamma/2}
    \end{bmatrix}
     e^{-i\Phi/2}
\label{eq: E_out}
\end{align}
By making use of the trigonometric identity
\begin{equation}
    e^{ia} + e^{-ia} = 2\cos(a) \,,
\end{equation}
we can rewrite Eq.~\ref{eq: E_out}, so that we end up with 
\begin{equation}
    E_{out} = \frac{E_0 \cos(\Gamma/2)}{\sqrt{2}}
    \begin{bmatrix}
        1 \\
        1 
    \end{bmatrix}
    e^{-i\Phi/2}\,.
\label{eq: E_out2} 
\end{equation}
\noindent Since we are interested in the intensity $I$ of the light detected by the camera, which is the squared norm of the Jones vector, $|E_{out}|^2 = \sum_i |E_{out,i}|^2$, we find that the measured intensity is
\begin{align}
        I_{meas} \propto  |E_{out}|^2 &= 2\times\frac{E_0^2 \cos^2(\Gamma/2)}{2} \\
        & = E_0^2 \cos^2(\Gamma/2)\,,
    \label{eq: I_meas} 
\end{align}
where the effective value of $\Gamma$ depends on the director tilt angle $\theta$. This is exactly the angle that is controlled by applying a voltage to the LCs, allowing us to modulate the outgoing amplitude and hence intensity of the light.
\pagebreak

\subsection{Experimental confirmation of the cos$^2$ response function}
\label{sec: experimental_confirmation_cos2}


To verify that the SLM follows the predicted $\cos^2$ behavior described in the previous section, we measure its response experimentally and fit the data to a theoretical model. The applied fit function is
\begin{equation}
\label{eq:fit_model}
    f(x) = A \cdot \cos^2(\frac{Bx}{2} + C) + D ,
\end{equation}
where $A$, $B$, $C$, and $D$ are free fit parameters. \newline

\begin{figure}[b!]
    \centering
    \includegraphics[width=1.0\linewidth]{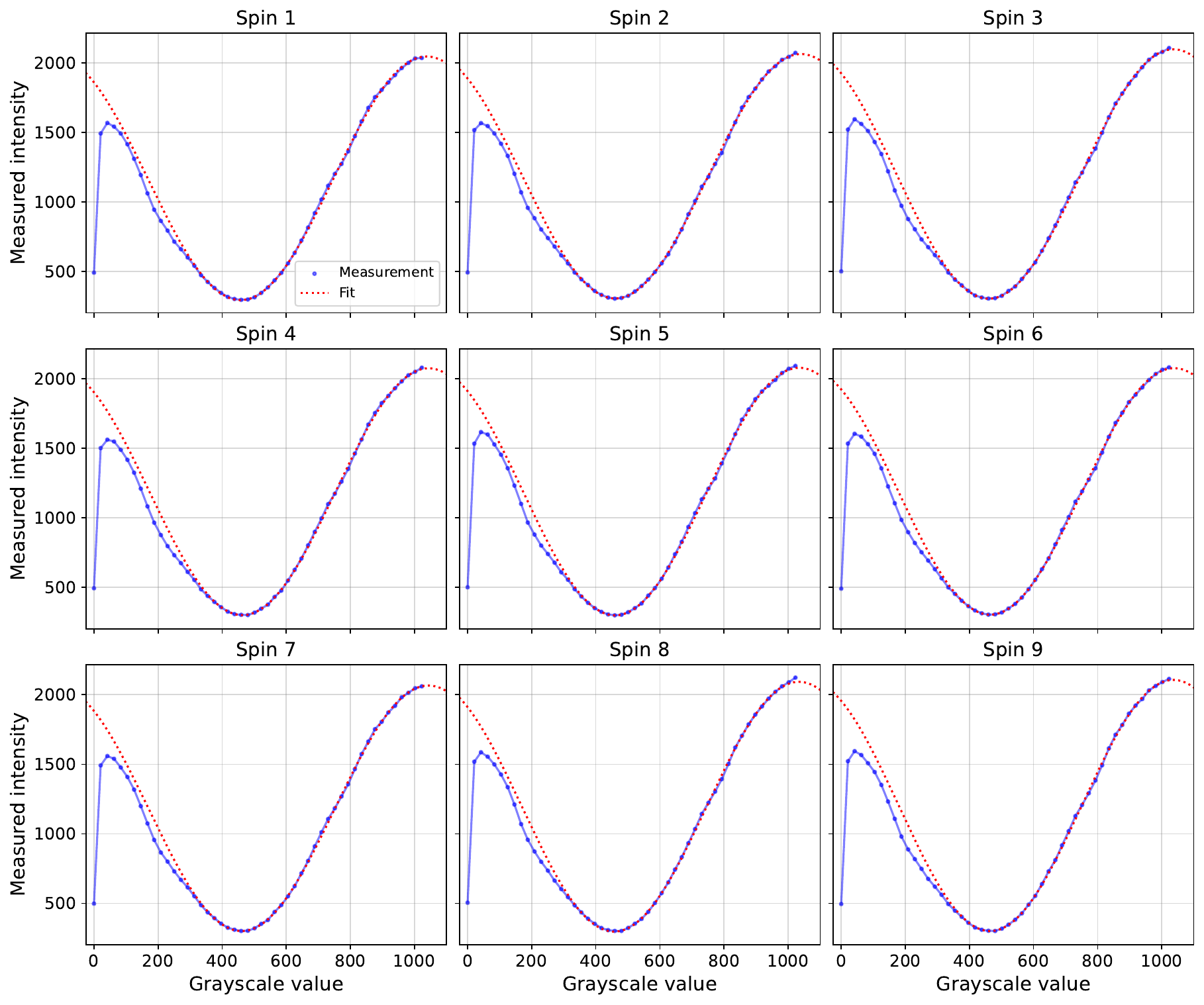}
    \caption{\textbf{Measured response functions of 9 spin variables together with their corresponding fits.} The 9 spins are arranged in a $3 \times 3$ lattice in the same order as shown in the figure. The blue dots represents the measured data, averaged over five runs, while the red dotted line shows the fitted model, obtained using only the data from the uphill slope.} 
    \label{fig:cos_transfer_example}
\end{figure}

To obtain the response function, we record the output intensity as a function of the discrete grayscale values applied to the SLM. Each grayscale value addresses a corresponding LC director angle $\theta$, defined as the angle between the LC director and the propagation direction, which sets the effective refractive index $n^{(1)}(\theta)$ and hence the phase retardation $\Gamma$ as described in Eq.~\ref{eq:Gamma_def}. The SLM provides 10-bit resolution, corresponding to 1024 discrete grayscale levels, calibrated by the manufacturer's look-up table to span a phase retardation range of $0$ to $2\pi$. We refer to these digitized drive values as the grayscale values. 
This resolution is high enough to implement the feedback-signal for most problems ~\cite{sevenants2025requirements}. Figure~\ref{fig:cos_transfer_example} shows the measured intensity as a function of the grayscale value applied to the SLM, for nine spin variables arranged in a $3\times3$ lattice. Since the intensities are recorded by the CMOS camera and subsequently digitized, they are reported in arbitrary units (a.u.). Each spin variable is encoded on a $4\times4$ pixel grid on the SLM with an additional 4 buffer pixels between 2 adjacent spins to avoid any influence between neighboring spins. The $3\times3$ lattice size was chosen to demonstrate the consistency of the response across spatially distributed spin variables, while keeping the presented fit data clearly legible. \newline

The blue markers represent the measured intensities, averaged over five independent runs. A nonlinear fit was then performed on the right-hand (uphill) half of the $\cos^2$ response function, using Eq.~\ref{eq:fit_model}. This restriction to the uphill slope is motivated by the unfaithful representation of the intrinsic $\cos^2$ response that can be seen in the downhill branch. At the lowest grayscale values, the LC molecules do not seem to complete their reorientation within the SLM's finite response time before the CMOS camera acquires the frame. Therefore, the initial drop in the recorded intensity at small grayscale values is an artifact of the SLM addressing dynamics rather than a feature of the underlying response function. Restricting the fit to the uphill slope, where this transient distortion is absent, therefore yields a more accurate estimate of the fit parameters.
\newline


The resulting fit parameters and their associated uncertainties, together with the $R^2$ and RMSE values of these fits, are summarized in Tab.~\ref{tab:spin_metrics}. The first two columns show consistently high $R^2$ values ($R^2 \geq 0.96$ for all nine spins) and comparatively small RMSE values. Expressed relative to the fitted amplitude $A$ (third column), the RMSE amounts to only $0.6$--$0.7\,\%$ of the signal range, meaning that the average deviation between the measured and fitted response function is more than two orders of magnitude smaller than the measured signal itself. Correspondingly, $R^2 \geq 0.96$ means the $\cos^2$ model accounts for at least $96\,\%$ of the variance in the measured response curve, leaving at most $4\,\%$ unexplained. Together, these values indicate excellent agreement between the experimentally measured response curves and the theoretical $\cos^2$ model derived in the previous section. Moreover, for every spin the fit uncertainty is small relative to the corresponding fitted value itself since the parenthetical one-standard-deviation uncertainty is one to two orders of magnitude smaller than the fitted number it accompanies, further confirming the reliability of the performed fit. \newline

During measurements, we also restrict the range over which the response function is used. This hard-clipping is motivated by three considerations. Firstly, the same asymmetry that led us to exclude the downhill slope from the fit makes the left-hand side of the curve less favorable for operation, so we again restrict ourselves to the uphill slope by clipping the response function at the minimum value of the $\cos^2$ response. Secondly, since the $\cos^2$ response is periodic, an unclipped system could in principle be driven by noise past the peak at large grayscale values in Fig.~\ref{fig:cos_transfer_example}. To avoid this, we hard-clip the response also at the maximum measured values of the uphill slope. Together with the lower cutoff, this produces a saturating, tanh-like response shape. The minimum and maximum values of the clipping are determined individually for each spin, based on the corresponding fits such as those shown in Fig.~\ref{fig:cos_transfer_example}, but typically lie close to grayscale values of 550 and 1000, respectively. Thirdly, this tanh-like shape is not only pragmatic. The tanh nonlinearity has been shown in literature to perform well for the class of IMs based on pitchfork bifurcation ~\cite{bohm2021order}.

\begin{table*}[b!]
    \centering
    \caption{\textbf{Fit parameters and goodness-of-fit metrics for the nine spins.} $A$, $B$, $\phi$ and $C$ are the parameters of the fit of Eq.~\ref{eq:fit_model}, while $R^2$ and the RMSE quantify the quality of that fit. Numbers in parentheses give one standard deviation in the last quoted digit(s). Fit is performed on the `uphill' slope of the cos$^2$.}
    \label{tab:spin_metrics}
    \footnotesize
    \setlength{\tabcolsep}{22pt}
    \begin{tabular}{
        c
        S[table-format=1.2]
        S[table-format=2.2]
        S[table-format=4.0(2)]
        S[table-format=1.3(1)]
        S[table-format=1.3(2)]
        S[table-format=3.0(1)]
    }
        \toprule
        Spin & {$R^2$} & {RMSE} & {$A$} & {$B$ [$10^{-3}$]} & {$\phi$ [rad]} & {$C$} \\
        \midrule
        1 & 0.97 & 12.46 & 1752(6)  & 5.413(4) & 0.332(14) & 293(2) \\
        2 & 0.98 & 10.99 & 1758(9)  & 5.421(4) & 0.322(12) & 304(1) \\
        3 & 0.97 & 11.89 & 1794(6)  & 5.450(3) & 0.316(11) & 303(1) \\
        4 & 0.98 & 10.53 & 1775(5)  & 5.411(4) & 0.317(11) & 300(1) \\
        5 & 0.96 & 13.27 & 1780(12) & 5.468(6) & 0.312(18) & 298(2) \\
        6 & 0.97 & 11.86 & 1773(8)  & 5.475(4) & 0.298(13) & 303(1) \\
        7 & 0.98 & 11.32 & 1766(7)  & 5.392(4) & 0.328(12) & 300(1) \\
        8 & 0.98 & 10.49 & 1796(8)  & 5.450(5) & 0.324(15) & 296(2) \\
        9 & 0.98 & 11.05 & 1806(10) & 5.496(5) & 0.293(14) & 300(1) \\
        \bottomrule
    \end{tabular}
\end{table*}

\pagebreak

\section{Calibration procedure steps}

In the previous section, we derived in detail how the cos$^2$ response is obtained from the used hardware as well as why we apply hard-clipping to it, resulting in the $\tanh$-like response function used throughout our measurements. Before an actual measurement can take place, however, the machine must first be calibrated. This initial calibration step is required to determine the position of the SLM pixels relative to the CMOS sensor, as well as the underlying response function of each individual spin. \newline

For an analog IM to operate correctly, each spin should be initialized at an intensity as close as possible to the center of its accessible intensity range, which minimizes any intrinsic bias toward the spin-up or spin-down state. We refer to this point as the bias point. Reaching this condition is more demanding in a spatially-multiplexed IM, since each spin must be biased individually, and this task becomes increasingly challenging as the number of spins grows. The reason is that the optimal initialization point can vary slightly from spin to spin, since spatial multiplexing inherently gives each spin its own physical implementation rather than a shared one, as is the case in time-multiplexed architectures. As a result, variations can arise in the nonlinear response of different spins, which can degrade computational performance if left uncorrected. In addition to being centered at the bias point, each spin must also be initialized with zero amplitude. This requires subtracting an offset intensity for every spin individually, so that its amplitude is truly zero when initialized at the bias point. \newline

\begin{figure}[b!]
    \centering
    \includegraphics[width=1.0\linewidth]{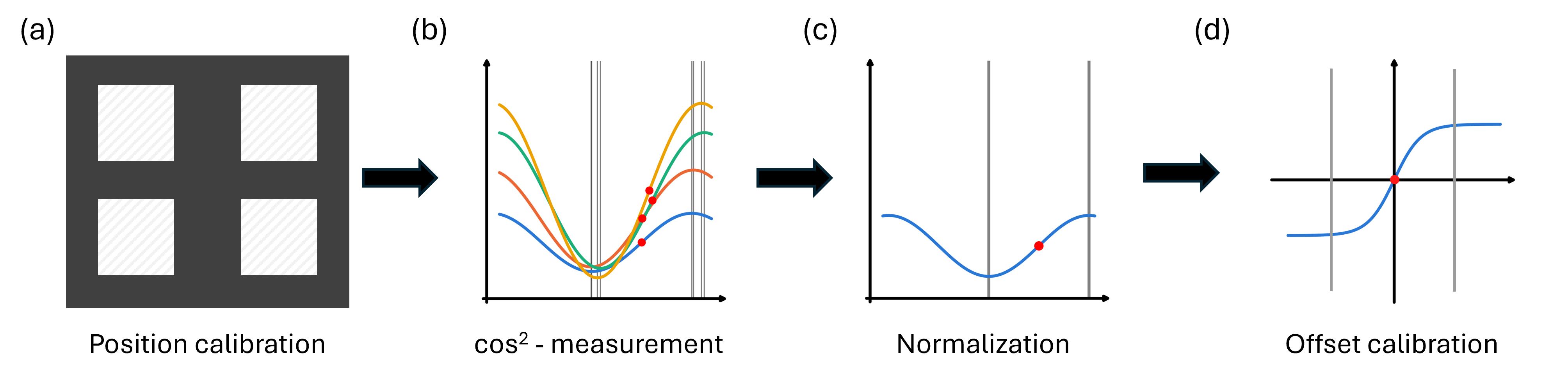}
    \caption{\textbf{Calibration sequence performed before each measurement.} \textbf{(a)} Position calibration: a high-contrast image of four spins (white, hatched) separated by dark buffer pixels is used to locate their positions accurately. \textbf{(b)} cos$^2$-response measurement: the intensity response of each spin is measured and fitted, giving its bias point (red dots) and clipping values (vertical lines). \textbf{(c)} Normalization: using the fit parameters from the previous step, the inhomogeneity between the different spins is removed. \textbf{(d)} Offset calibration: each spin is initialized at its bias point and the corresponding offset is subtracted, centering its amplitude at zero. Combined with the clipping bounds, this yields the tanh-like response function used during operation.}
    \label{fig:calibration_steps}
\end{figure}

To address these challenges, we developed a fully autonomous calibration script that carries out the entire calibration procedure without any user intervention. Figure~\ref{fig:calibration_steps} shows a schematic illustration of these four steps and each of them is explained in more detail here below.

\subsection{Position calibration}

In our ASIM setup, individual spin variables are created by assigning each spin to a distinct region of the SLM and imaging these regions onto separate areas of a CMOS camera sensor. The spatial coordinates of the spins defined on the SLM cannot, however, be mapped directly onto the CMOS detector. This mismatch arises both from the difference in resolution between the SLM and the camera, and from imperfections in the imaging system, which produce a gradual intensity distribution at the spin boundaries rather than a sharp cutoff. Before any measurement can begin, the positions of the spins on the CMOS camera must therefore first be determined.\newline

Position calibration compensates for the spatial mismatch between the SLM and CMOS pixels by establishing the correspondence between the two. A series of calibration patterns is projected onto the SLM and recorded by the CMOS camera, and threshold-based segmentation is used to identify the high-intensity regions corresponding to the activated spins. The centroids of these regions are used to fit a regular grid describing the spin positions on the camera. Each grid position is subsequently associated with its corresponding image region, defining the set of camera pixels that represents each spin and is used for all subsequent measurements throughout the experiment.

\subsection{cos$^2$-calibration}

Once the position of each spin has been identified, the algorithm performs a response-function calibration for every spin individually. This is done by sweeping the phase applied to the SLM and recording the resulting intensity on the CMOS camera, a procedure already illustrated in section~\ref{sec: experimental_confirmation_cos2}. As shown there, the response function of each spin can be reconstructed from these measurements and fitted to the expected cos$^2$ behavior.\newline

This procedure serves several purposes. First, it allows us to assess the quality of the nonlinear response across all spins by calculating the corresponding $R^2$- and RMSE-values. Second, it allows us to determine the points at which the cos$^2$ will later be clipped. Finally, it enables the extraction of key parameters of each response function, namely the amplitude of the cos$^2$ and the global intensity offset, which prevents the minimum of the cos$^2$ from lying at zero. \newline

Together, these parameters quantify the variation in response functions between spins, meaning that different spins can react differently even when driven by the same feedback signal. This is undesirable, as it can degrade performance and ideally, every spin should respond identically to the same input. We attribute these differences (in amplitude and global intensity offset) primarily to the non-uniform illumination of the LED, and expect that a more uniform light bundle would already reduce the spread between spins considerably. Importantly, despite these differences in fit parameters, all spins follow the same functional form, which allows us to apply a normalization step that removes as much of this variation as possible.

\subsection{Normalization}

To compensate for the differences between the individual cos$^2$ response functions, the algorithm uses the fitted parameters from the previous calibration step to normalize the measured response functions between 0 and 1. Specifically, the global intensity offset found for each spin is used to shift its minimum to zero, while dividing by the corresponding amplitude fit-parameter scales its maximum to 1. These normalized values are then used when computing the feedback signal, ensuring that each spin receives the appropriate amount of feedback so that all spins behave in the same way. In effect, this normalization step produces a set of far more homogeneous cos$^2$ response functions. Note that this step would become unnecessary if a uniform illumination beam were used instead.

\subsection{Offset calibration}

The cos$^2$-calibration step also provides the bias point of each spin, i.e. the center of its accessible intensity range. This bias point is an important parameter as it minimizes the bias toward the spin-up or spin-down state. For the cos$^2$ response in our setup, this corresponds to the inflection point, specifically that on the "uphill" slope, as discussed in section~\ref{sec: experimental_confirmation_cos2}. This bias point is the phase-value at which each spin should be initialized. \newline

Once the algorithm initializes the spins at their bias point, an offset calibration is performed by measuring the intensity for each spin over several iterations. Subtracting this offset intensity ensures that every spin starts from an unbiased state with zero amplitude. Due to the digitally addressable nature of the ASIM, this procedure can be applied individually and straightforwardly to each spin variable. Combined with the hard-clipping introduced earlier, this results in the $\tanh$-like response function, centered around zero, that we use throughout our measurements.

\pagebreak

\section{Experimental scaling of the $R^2$ and RMSE-values}

\subsection{CMOS camera settings} 
The CMOS camera used (FLIR Grasshopper, GS3-U3-32S4M) has a 12-bit pixel depth and two key adjustable parameters, the gain and the acquisition time, both of which strongly influence the effective dynamic range and overall measurement quality. \newline

The CMOS gain corresponds to the electronic amplification applied to the pixel signal prior to analog-to-digital conversion. Increasing the gain enhances weak signals and can help mitigate the relative influence of readout noise. However, because the noise contributed by the amplification circuitry itself increases with gain, increasing gain generally reduces the SNR. In contrast, the acquisition time (also known as the integration time) defines the duration over which incident photons are collected. Longer acquisition times increase the detected photon number, typically improving the SNR and extending the usable dynamic range, provided that saturation is avoided. \newline

To evaluate the influence of these two parameters on our experimental setup, we assess the quality of the measured response function based on measurements such as those shown in Fig.~\ref{fig:cos_transfer_example}. The measured response is fitted with the theoretically expected $\cos^2$ dependence using only the `uphill' part, and the goodness of fit is quantified using the coefficient of determination ($R^2$) and the root-mean-square error (RMSE). While $R^2$ measures how well the model explains the variance in the data, the RMSE provides an absolute measure of the fitting error. \newline

\begin{figure}[t!]
    \centering
    \includegraphics[width=1\linewidth]{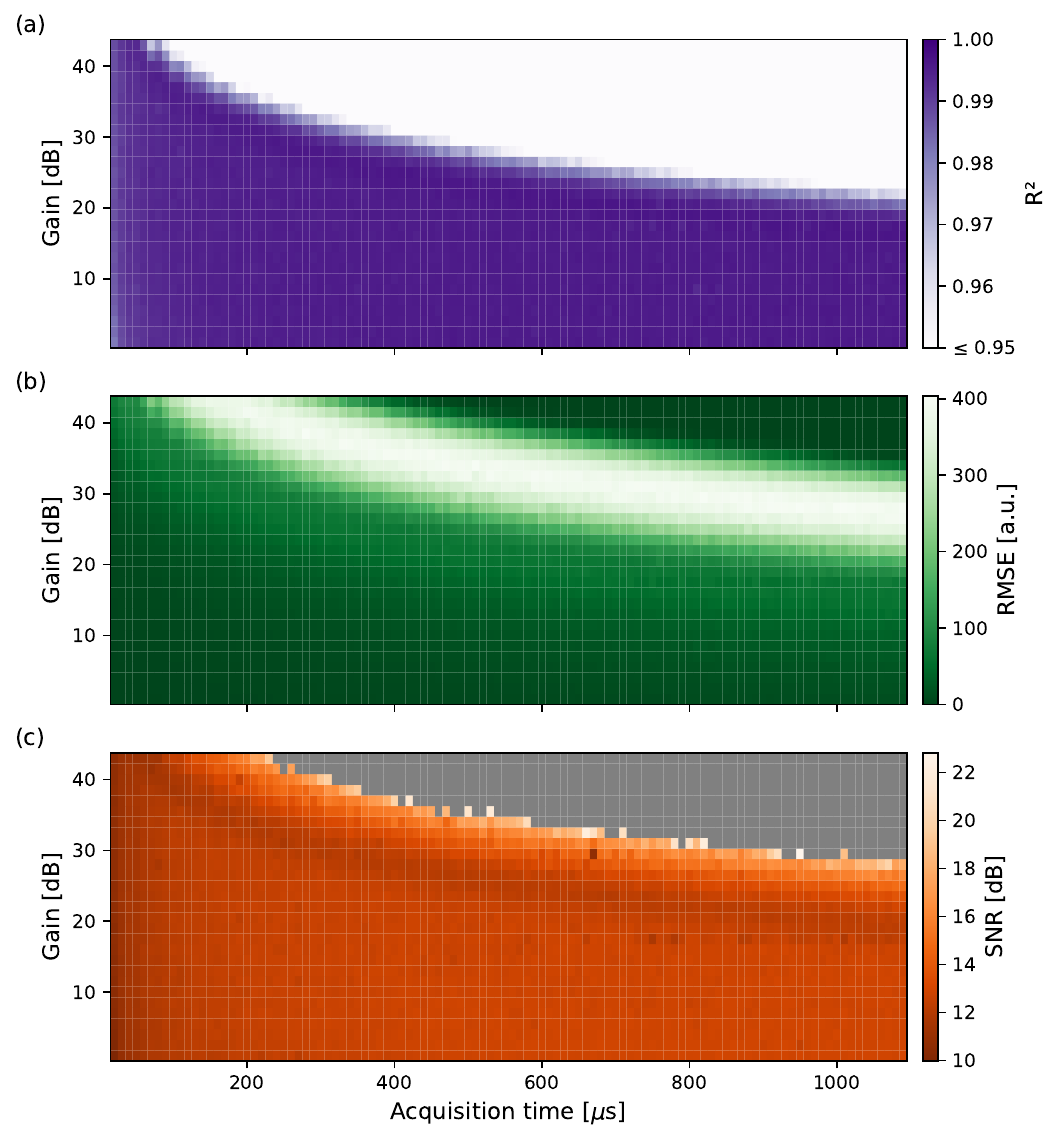}
    \caption{\textbf{$\boldsymbol{R^2}$, RMSE, and SNR of the $\boldsymbol{\cos^2}$ fit as a function of CMOS gain and acquisition time.}
    \textbf{(a)} $R^2$ value of the $\cos^2$ fit for different gain and acquisition time combinations. The lower limit of the color scale is truncated to emphasize variations in the high $R^2$ regime. Higher values indicate better agreement with the theoretical expected model.
    \textbf{(b)} RMSE of the corresponding fits. lower values indicate a better fit quality (lower error).
    \textbf{(c)} SNR measured at the bias point for the same gain and acquisition time settings of the CMOS camera. Gray cells indicate undefined SNR values, caused by a zero $\sigma_{\mathrm{bias}}$ resulting from signal clipping.
}
    \label{fig:cmos_settings_scan}
\end{figure}

Fig.~\ref{fig:cmos_settings_scan}(a) and (b) show the resulting $R^2$ and RMSE values, respectively, for a systematic scan over gain and acquisition time. The gain was varied from 1\,dB to 43\,dB (in steps of 1.5\,dB), and the acquisition time from 20\,$\mu$s to 1100\,$\mu$s (in steps of 10\,$\mu$s). Optimal CMOS settings correspond to parameter combinations that simultaneously yield a high $R^2$ value and a low RMSE. \newline

In Fig.~\ref{fig:cmos_settings_scan}(a), two distinct regions can be identified. For low to moderate gain and acquisition time (bottom region), the fit quality is excellent, with $R^2 > 0.99$. In contrast, for large gain and long acquisition times (upper-right region), the $R^2$ value drops below 0.95. This degradation is caused by signal saturation. When the detector approaches its maximum digital value, the recorded signal no longer follows the expected $\cos^2$ modulation. The resulting distortion reduces the agreement with the theoretical model. For visualization purposes, the lower limit of the color scale is set to 0.95 to enhance contrast in the high-$R^2$ regime but the actual $R^2$ values in the saturated region are significantly lower. \newline

The RMSE map in Fig.~\ref{fig:cmos_settings_scan}(b) reveals three characteristic regions. In the lower-left region (low gain and short acquisition time), the RMSE is small, reflecting the excellent agreement between measurement and model. As gain and acquisition time increase, saturation effects begin to distort the signal, leading to an increase in RMSE. For very high gain and long acquisition times, the detector output becomes fully saturated, resulting in an almost constant signal at the maximum digital level. In this regime, the data exhibit little variation. Consequently, the nonlinear fitting procedure converges toward a nearly flat solution with minimal residual fluctuations. Although this can produce artificially low RMSE values, the fit no longer captures the underlying physical $\cos^2$ behavior. This situation corresponds to a poor explanatory power of the model, as reflected by the very low $R^2$ values observed in the same parameter region in Fig.~\ref{fig:cmos_settings_scan}(a). Statistically, the reason is that the model suffers from a loss of variance in the data, making the fit formally small in residuals but physically meaningless. \newline

Combining both metrics, the optimal CMOS settings correspond to the broad lower region where $R^2$ remains close to unity and the RMSE is minimal. This indicates that a relatively large parameter space yields reliable measurements. To further refine the choice of operating point within this region, we also consider the signal-to-noise ratio (SNR). \newline

We define the SNR at the inflection point of the uphill slope of the $\cos^2$ response function. This point is chosen as the bias point of the IM. Physically, it corresponds to the grayscale value at which the IM is initiated. After subtracting the background offset, this operating point is the zero-amplitude reference of the spin variables. The SNR is evaluated at this bias point because noise contributions are relatively more significant when the spin amplitudes are small as it can easily flip the spins. In contrast, at saturated amplitudes, the noise has a harder time influencing the spin dynamics because the feedback signal is already dominated by the gain and mutual-coupling term, making the impact of noise less critical. Therefore, characterizing the SNR at the inflection point provides a more relevant measure of performance in the regime where the system is most sensitive to noise. \newline

Once initiated at the bias point, the intensity at that point is measured for 80 iterations. Thereafter, we calculate the mean (I$_{bias}$) and the standard deviation ($\sigma_{bias}$) of this time-series. The SNR is then calculated as  
\begin{equation}
    SNR_{dB} = 10 \log_{10} \left( \frac{I_{bias}}{\sigma_{bias}} \right),
\label{eq:SNR_def}
\end{equation}
which is expressed in decibel (dB). The resulting SNR values are shown in Fig.~\ref{fig:cmos_settings_scan}(c). The SNR ranges from approximately 10\,dB up to 20\,dB. However, regions with higher SNR values also overlap with those of higher error in the response function, as seen in Fig.~\ref{fig:cmos_settings_scan}(b), indicating that a high SNR alone is not a reliable indicator of fit quality. The gray region at the top of Fig.~\ref{fig:cmos_settings_scan}(c) corresponds to undefined SNR values, which occur when $\sigma_{\mathrm{bias}} = 0$ as a consequence of the signal clipping discussed earlier. These settings are therefore unusable. Taken together, these observations show that there is a wide operating regime for the CMOS camera  
where the SNR remains adequate, the $\cos^2$ response is fitted with low error, and the signal stays unclipped. Within the scanned range, this corresponds to a CMOS gain $\in$ [2, 20]\,dB and a acquisition time $\in$ [40, 1100]\,$\mu$s. 
The exact values that were used during the measurements are reported in the next subsection.

\subsection{The $R^2$ and RMSE-values scaling with problem size}

To confirm that the quality of the calibration procedure of the spins remains robust over a wide range of problem sizes, we track how the average $R^2$ value and the normalized root-mean-square error (NRMSE) of the fitted response function scale with problem size $N$. The NRMSE is the RMSE of the fit, normalized with respect to the fitted amplitude. Since both quantities carry the same unit, this ratio is dimensionless and can be compared directly across problem sizes. \newline
 
\begin{figure}[t!]
    \centering
    \includegraphics[width=\linewidth]{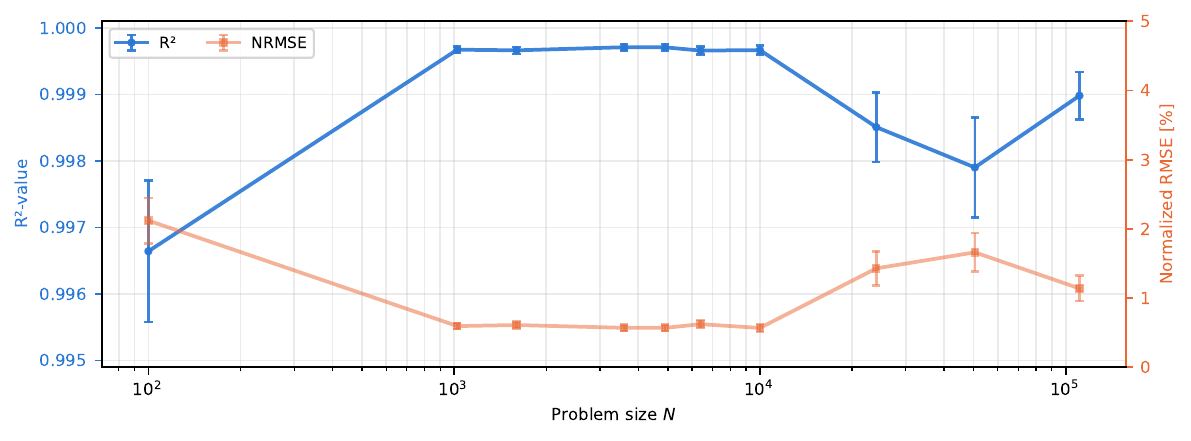}
    \caption{\textbf{Calibration metrics for the spins.} The average $R^2$ (blue) and NRMSE (orange) values are shown as a function of problem size $N$, both averaged over all spins in the lattice. Error bars indicate one standard deviation of the corresponding distribution. All problem sizes satisfy $R^2 > 0.995$ and NRMSE $< 2.5\%$, confirming that the calibration quality of the spins remains high across the full range of problem sizes considered.}
    \label{fig:RMSE_R2_scaling}
\end{figure}

\begin{table*}[b!]
    \centering
    \caption{\textbf{Spin and CMOS acquisition parameters for each problem size.} For every antiferromagnetic square-lattice problem of size $N$, the table lists the spin size and spacing programmed on the SLM, together with the corresponding CMOS acquisition time and gain.}
    \label{tab:spin_settings}
    \footnotesize
    \begin{tabular}{>{\raggedright\arraybackslash}p{2.3cm} r c c c c}
        \toprule
        Problem & $N$ & Spin size [pixels]$^\dag$ & Spacing [pixels]$^\ddag$ & CMOS acquisition time [\textmu s] & CMOS gain [dB] \\
        \midrule
        \multicolumn{6}{l}{\textit{Square lattice}} \\
         & 100      & $4\times4$ & 4 & 45 & 15 \\
         & 1024     & $4\times4$ & 4 & 45 & 15 \\
         & 1600     & $4\times4$ & 4 & 45 & 15 \\
         & 3600     & $4\times4$ & 4 & 45 & 15 \\
         & 4900     & $4\times4$ & 4 & 45 & 15 \\
         & 6400     & $4\times4$ & 4 & 45 & 15 \\
         & 10{,}000  & $3 \times 3$ & 3 & 45 & 15 \\
         & 24{,}025  & $1 \times 1$ & 2 & 1006 & 10 \\
         & 50{,}451  & $1 \times 1$ & 2 & 1006 & 10\\
         & 110{,}889 & $1 \times 1$ & 1 & 1006 & 10 \\
        \bottomrule
        \addlinespace[2pt]
        \multicolumn{6}{@{}p{\textwidth}@{}}{\footnotesize $^{\dag}$\,Spin size denotes the $n \times n$ cluster of SLM pixels used to image a single spin onto the CMOS camera $^{\ddag}$\, Spacing denotes the number of buffer pixels between two adjacent spins to avoid electrical cross-talk between neighbouring SLM pixels and optical cross-talk from diffraction on the CMOS sensor.}
    \end{tabular}
\end{table*}

Fig.~\ref{fig:RMSE_R2_scaling} shows the average $R^2$ and normalized RMSE (NRMSE) values, averaged over all spins in the lattice, as a function of problem size. The error bars represent one standard deviation of the underlying distribution of $R^2$ and NRMSE values of the individual spins, rather than a measurement uncertainty on the mean itself. Therefore, this spread should be interpreted as how much the fit quality varies from spin to spin at a given problem size. For each spin, we repeatedly measured the response function, averaged these repeated measurements, and fitted Eq.~\ref{eq:fit_model} to the resulting averages, weighted by their standard error. The data shows a dip for problem sizes between $1{,}000$ and $10{,}000$. This is attributed to the fact that we used ten repeated measurements to calibrate the spins in this range, compared to three repeated measurements for the smallest problem size (100 spins) and five for the largest problem sizes ($>10{,}000$ spins). The smallest problem size was the first calibration performed, before we had settled on a repeat count for the remaining problem sizes. We then used ten repeats over the intermediate range to more rigorously confirm that the calibration procedure continued to hold up as problem size increased, and reduced this to five repeats for the largest sizes to save time without compromising fit quality. Due to the increase in $R^2$ and decrease in RMSE with more repeated measurements, Fig.~\ref{fig:RMSE_R2_scaling} also confirms that averaging over more repeats results in better-quality fits. Nonetheless, all spins have good fit metrics regardless. This is firstly seen from the $R^2$, which is consistently higher than 0.995, but also from the NRMSE, which is lower than 2.5\% for every problem size considered. This shows that our setup achieves high-quality response functions even for the largest sizes considered in this work. Note that the higher $R^2$ values reported here, compared to those in Tab.\ref{tab:spin_metrics}, result from the fact that the fit was carried out using the averaged response functions. \newline

Reaching the largest problem sizes considered in this work required continuously adapting the physical implementation of the spins as the setup was scaled up, in particular by decreasing the physical size of the spins. Each spin is implemented physically as a cluster of $n \times n$ SLM pixels, where the size of this cluster has varied over the course of the setup’s development. Since the spin size plays an important role in both the achievable problem size and the calibration quality, we report the corresponding settings in Tab.~\ref{tab:spin_settings}, where this cluster size is listed as the spin size. Between neighbouring spin clusters we also leave a margin of unused pixels, reported as the spacing. This spacing is needed to suppress electrical cross-talk between adjacent SLM pixels as well as optical cross-talk caused by diffraction on the CMOS sensor. For each problem size, Tab.~\ref{tab:spin_settings} lists the resulting spin size and spacing in pixels, together with the acquisition time and gain of the CMOS camera used to read out the corresponding intensity pattern. These settings were adjusted for each problem size to ensure every spin was bright enough for the position calibration step, which locates each spin in the camera image, to reliably detect it.

\pagebreak

\section{Long-term stability of the spin amplitude}

Hardware implementations of an analog IM store the spin state in a physical quantity, and the fidelity of the computation is bounded by the stability of that quantity. In the ASIM, each spin variable is encoded in the optical intensity so a slow drift of the measured intensity would be fed back into the dynamics, generating a bias for the spin-variables. Long-term stability is therefore essential for reliable performance. \newline

\begin{figure}[b!]
    \centering
    \includegraphics[width=\linewidth]{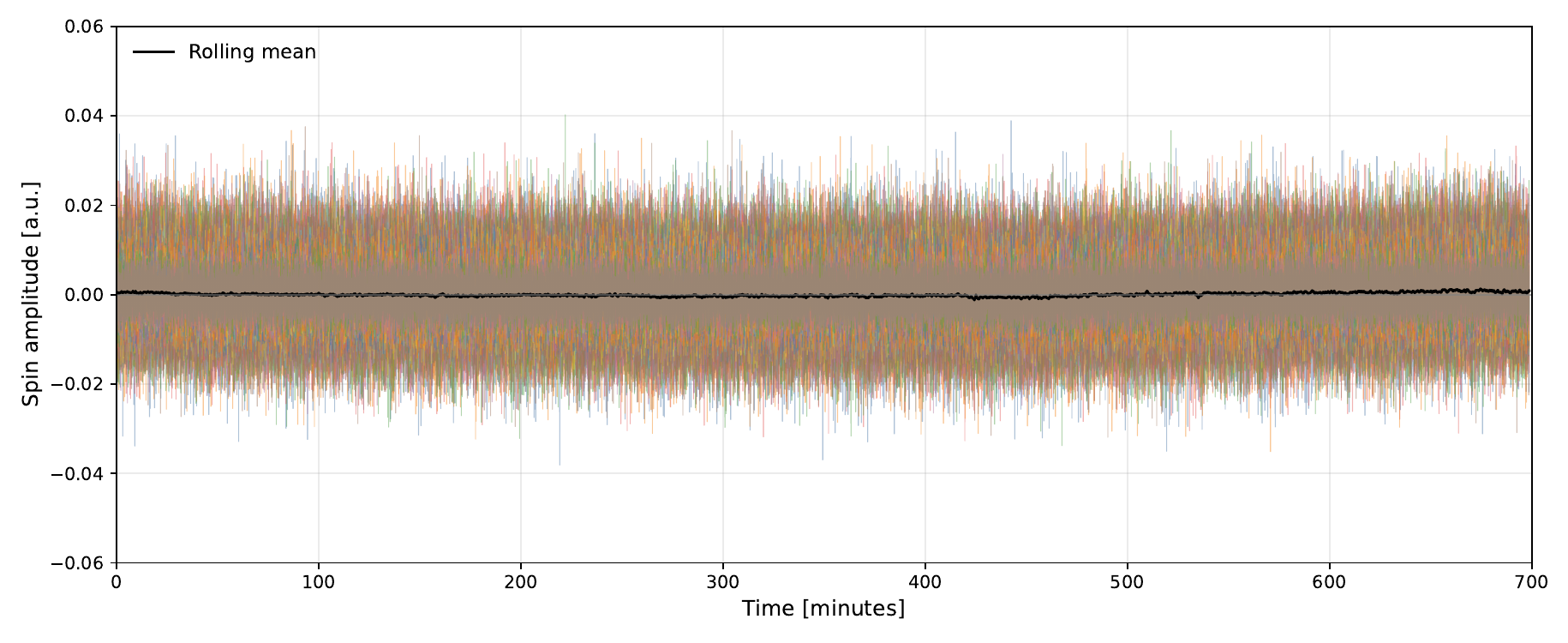}
    \caption{\textbf{Long-term stability of the spin-readout amplitude.} The measured analog spin amplitudes for 100 uncoupled spins ($\alpha = \beta = 0$) are shown in colored lines. A centered rolling mean with a window $\Delta t$ of $42$s is added in black to suppresses the shot-to-shot fluctuations and visualizes the behavior of the mean signal. We observe no systematic drift over the $\sim$700~minutes of continuous operation.
    }
    \label{fig:warmup_stability}
\end{figure}

Two properties of our architecture work in our favor. First, the information is carried by the intensity of the field rather than by its phase, so the optical path-length changes produced by thermal expansion or air currents, which would drastically change the phase in a coherent Ising machine, leave our measured intensity unaffected. The system can therefore operate at room temperature with no temperature stabilization, no enclosure, and no active feedback on the optical path. Second, the only sensitivity that remains in our setup is the relative position of the SLM and the CMOS camera. Since the pixels of both devices are of micrometer scale, the two planes must stay in a stable position with respect to each other, which we achieve passively using a pneumatically isolated optical table that suppresses high-frequency vibrations. Residual low-frequency motion is not guaranteed to be suppressed by this isolation, but we resolve no signature of it in the data below. \newline

Figure~\ref{fig:warmup_stability} characterizes the long-term stability of the spin-readout amplitude over a continuous $\sim$700-minute acquisition following the initial system warm-up. In Fig.~\ref{fig:warmup_stability}, the spin amplitude for each spin is plotted as well as a moving average over all spins with a window size of $\Delta t = 42$s (black line). The moving average suppresses the shot-to-shot fluctuations and visualizes the behavior of the mean signal. We observe that the rolling mean remains relatively flat and consistently within the noise envelope of the individual traces, indicating that no systematic drift, such as a residual warm-up effect, is present over the full measurement window.

\pagebreak

\section{Benchmarking details}

Throughout the benchmarking procedure, we employed two distinct strategies to enhance the probability of attaining the optimal solution. For sparse problem instances, we applied a linear annealing schedule to the $\beta$ parameter, increasing $\beta$ from an initial value to a final value over the course of the measurement process. For denser problem instances, we introduced a mixing parameter $h$ to simultaneously reduce the sensitivity to hyperparameter selection and to increase the probability of obtaining the ground-state solution. The precise definitions and implementation details of these strategies are presented in the subsections below. Finally, given that the performance of any heuristic solver depends critically on the choice of parameters, we additionally report the parameter settings used, together with other relevant performance metrics, in the third and final subsection of this section.

\subsection{Linear annealing schedule}

Running an IM with a fixed pair of hyperparameters $(\alpha, \beta)$ throughout the evolution has the drawback that from the moment the system is initialized, the energy landscape it must navigate contains many local minima, which lowers the probability that the IM will converge to the ground-state solution. Annealing schedules are often used in IMs to improve performance this, and can be applied to the noise level, the feedback strength $\alpha$, or the spin-spin coupling strength $\beta$ ~\cite{pierangeli2020noise, yamamura2024geometric, pierangeli2020adiabatic}. In this work we anneal $\beta$, which controls how strongly the spin network couples into the solver's dynamics. The expected improvement in performance is based on the insight that immediately after the pitchfork-bifurcation, there is only one stable fixed point. \newline

At $\beta = 0$ the spins are uncoupled and if $\alpha < 1$ the system has only the trivial all-zero solution, meaning the energy landscape is convex, with a single fixed point. As $\beta$ is increased, this landscape gradually deforms into the full, rugged Ising energy landscape whose global minimum encodes the optimal solution. Ramping $\beta$ up smoothly therefore lets the problem's influence on the dynamics grow gradually, exposing the IM to the rugged landscape step by step rather than all at once, which in turn makes it easier for the IM to find its way to the ground state. \newline

Concretely, we use a linear schedule that increases $\beta$ at every iteration $k$ from an initial value $\beta_{start}$ to a final value $\beta_{end}$:
\begin{equation}
    \beta[k] = \beta_{start} + \frac{k}{N_{iter}}(\beta_{end} - \beta_{start}) \,,
\end{equation}
where $\beta_{end}$, as well as $\alpha$, is determined from brute-force parameter simulations and $N_{iter}$ is the number of iterations of a single run. $\beta_{start}$ is calculated using
\begin{equation}
\label{eq: bifurcation_point}
    \beta_{start} = \frac{1-\alpha}{\mu_{+}} \,,
\end{equation}
where $\mu_+$ is the largest eigenvalue of the coupling matrix \textbf{J}. Together with $\alpha$, it corresponds to the bifurcation point, the point at which the trivial zero-amplitude fixed point becomes unstable and only one stable fixed point exists. Starting the annealing here avoids two opposite pitfalls. Starting at a lower $\beta$ would be wasteful, since the IM cannot bifurcate below this point and therefore cannot reach the ground state of the encoded problem, wasting all the iterations spend in that regime. Starting at a higher $\beta$, on the other hand, would launch the IM directly into an energy landscape that already contains many local minima, which is exactly what annealing is meant to avoid. Analytically calculating the bifurcation point using Eq.~\ref{eq: bifurcation_point} therefore gives an ideal choice for $\beta_{start}$. A more detailed derivation of Eq.~\ref{eq: bifurcation_point} is given in Ref.~\cite{de2025incorporating}.

\subsection{Mixing parameter}
 
For denser problems, we empirically found that adding a mixing parameter improves the performance of our solver. The idea is taken from Ref.~\cite{sevenants2026reducing}, where it is shown that measurement-feedback-based pitchfork bifurcation systems with a mixing term show lower sensitivity to the choice of hyperparameters. We use it here because we empirically found it hard to identify suitable hyperparameters for the denser problem instances, which include the number-partitioning problems and the BiqMac $g05\_80.1$ instance, with edge densities of 1 and 0.5 respectively. \newline
 
The mixing strategy is based on the following update rule:
\begin{equation}
    x_i[k+1] = (1-h)\, x_i[k] + h\, \tanh\!\left(f_i + \gamma \zeta_i[k]\right),
\end{equation}
where $x_i[k]$ is the amplitude of spin variable $i$ at iteration $k$, $f_i$ is the feedback term for spin $i$ defined in the main text, and $\gamma$ controls the strength of the Gaussian white noise $\zeta_i$ ($\mu = 0$, $\sigma^2 = 1$). \newline
 
The mixing strategy thus combines the amplitude from the previous iteration with the feedback signal, weighted by a mixing term $h \le 1.0$. Note that for $h = 1.0$, the previous-iteration term drops out and we recover the unmodified, time-discrete dynamics of our original setup, which are driven purely by the feedback signal. Since this feedback is calculated digitally, introducing the mixing term $h$ is only a minor modification to the code. Importantly, the mixing term does not increase the time-to-solution, and in the best case even decreases it \cite{sevenants2026reducing}. \newline

\subsection{Experimental details and performance metrics}

\begin{table*}[t!]
    \centering
    \caption{Hyperparameters and performance metrics for all benchmark problem instances. The solution quality (SQ) is the best solution found within a run relative to the best-known solution, the success rate (SR) is the fraction of runs reaching the best-known solution, $SR_{99}$\,$^{\dag}$ and $SR_{95}$\,$^{\dag}$ are the fractions of runs reaching, respectively, 99\% and 95\% of the best-known solution, and the ITS is the number of iterations required to reach the best-known solution with 99\% probability. NA indicates a parameter that was not used for that benchmark, while a cross ($\times$) indicates a metric that could not be calculated for that instance.}
    \label{tab:hyperparameters}
    \footnotesize
    \begin{tabular}{>{\raggedright\arraybackslash}p{1.5cm} r c c c c c c c c c c c}
        \toprule
        Problem & $N$ & $\alpha$ & $\beta_{start}$ & $\beta_{end}$ & $\gamma$ & $h$ & Iterations & SQ [\%] & SR [\%] & SR$_{99}$[\%] & SR$_{95}$ [\%] & ITS [iterations] \\
        \midrule
        \multicolumn{13}{l}{\textit{Square lattice}} \\
         & 100   & 0.5555 & 0.1111 & 0.4444 & 0.001 & NA & 500 & 100.00 & 98 & 98 & 98 & 230 \\
         & 1024  & 0.4444 & 0.1389 & 0.4444 & 0.001 & NA & 1000 & 100.00 & 100 & 100 & 100 & 840 \\
         & 1600  & 0.2222 & 0.1945 & 0.5555 & 0.001 & NA & 1500 & 100.00 & 100 & 100 & 100 & 834 \\
         & 3600  & 0.1111 & 0.2222 & 0.6666 & 0.001 & NA & 3000 & 100.00 & 100 & 100 & 100 & 1763 \\
         & 4900  & 0.3333 & 0.1667 & 1.0000 & 0.001 & NA & 3500 & 100.00 & 84 & 84 & 98 & 4706 \\
         & 6400  & 0.3333 & 0.1667 & 0.4444 & 0.001 & NA & 4500 & 100.00 & 86 & 86 & 100 & 10 324 \\
         & 10\,000 & 0.3333 & 0.1667 & 0.4444 & 0.001 & NA & 6000 & 100.00 & 24 & 56 & 100 & 100 347 \\
         & 24\,025 & 0.3666 & 0.1584 & 1.0000 & 0.001 & NA & 3000 & 98.56 & 0 & 0 & 86 & $\times$ \\
         & 50\,451 & 0.3333 & 0.1667 & 0.8888 & 0.001 & NA & 6000 & 96.49 & 0 & 0 & 14 & $\times$ \\
         & 110\,889 & 0.3333 & 0.1667 & 0.5555 & 0.001 & NA & 10000 & 85.72 & 0 & 0 & 0 & $\times$ \\
        \midrule
        \multicolumn{13}{l}{\textit{G-set}} \\
        G1 & 800 & 0.95 & 0.0785 & 0.0038 & 0.001 & NA & 1500 & 99.43 & 0 & 90 & 100 & $\times$ \\
        \midrule
        \multicolumn{13}{l}{\textit{BiqMac}} \\
        g05\_80.1 & 80 & 0.9 & 0.04 & NA & 0.01 & 0.5 & 500 & 100 & 84 & 98 & 100 & 700 \\
        \midrule
        \multicolumn{13}{l}{\textit{Number partitioning}} \\
         & 32   & 0.60 & 5e-4 & NA & 0.05 & 0.1 & 100 & 100 & 42 & NA 
         & NA 
         & 227 \\
         & 64   & 0.40 & 1e-4 & NA & 0.05 & 0.1 & 100 & 100 & 24 & NA 
         & NA 
         & 1579 \\
         & 256  & 0.65 & 5e-5 & NA & 0.05 & 0.1 & 150 & 100 & 32 & NA 
         & NA 
         & 1030 \\
         & 768  & 0.80 & 1e-6 & NA & 0.05 & 0.1 & 500 & 100 & 40 & NA 
         & NA 
         & 3916 \\
         & 1280 & 0.70 & 5e-7 & NA & 0.05 & 0.1 & 800 & 100 & 22 & NA 
         & NA 
         & 9521  \\
        \bottomrule
        \addlinespace[2pt]
        \multicolumn{13}{@{}p{\textwidth}@{}}{\footnotesize $^{\dag}$\,$SR_{99}$ and $SR_{95}$ are evaluated on the energy for the square-lattice instances and on the cut value for the G-set and BiqMac instances.}
    \end{tabular}
\end{table*}

In this subsection we give a detailed overview of the hyperparameters used throughout our experiments, together with the metrics used to evaluate the performance of the ASIM. All hyperparameters and performance metrics for the individual benchmark problems are summarized in Tab.~\ref{tab:hyperparameters}. \newline

The reported hyperparameters are the feedback strength $\alpha$, the initial and final spin-spin coupling strengths $\beta_{start}$ and $\beta_{end}$, the digitally added noise strength $\gamma$, the mixing parameter $h$ and the number of iterations per run. As discussed above, the annealing scheme was not used for the denser problems. For these instances we mark the $\beta_{end}$ column as non-applicable (NA), indicating that $\beta$ was instead kept fixed at $\beta_{start}$ throughout the run. To quantify the performance of the ASIM we report three metrics: the solution quality (SQ), the success rate (SR) and the iterations-to-solution (ITS), each computed over $N_{\mathrm{runs}} = 50$ independent runs per problem instance, except for the antiferromagnetic square lattice with problem size $> 10 000$ spins. The solution quality is the best solution found within a run, expressed as a percentage of the best-known solution for that instance. The success rate is the probability that a single run reaches the best-known solution, either by converging or transiently, and is calculated as
\begin{equation}
    \label{eq:SR_def}
    SR = \frac{N_{\mathrm{success}}}{N_{\mathrm{runs}}},
\end{equation}
where $N_{\mathrm{success}}$ is the number of runs in which the best-known solution is reached. For the square lattice, the G-set and the BiqMac instances, Table~\ref{tab:hyperparameters} additionally lists $SR_{99}$ and $SR_{95}$ defined identically but with the target value relaxed to 99\% and 95\% of the best-known solution. $SR_{99}$ and $SR_{95}$ are evaluated on the energy for the square-lattice instances and on the cut value for the G-set and BiqMac instances. \newline

The iterations-to-solution is the number of iterations required to reach the best-known solution with 99\% probability~\cite{im2026accelerating},
\begin{equation}
    \label{eq:ITS_def}
    \mathrm{ITS}(N_{\mathrm{iter}})
    =
    N_{\mathrm{iter}}
    \max\!\left\{
        1,\;
        \frac{\ln(1-0.99)}{\ln\bigl(1-SR(N_{\mathrm{iter}})\bigr)}
    \right\},
\end{equation}
where $N_{\mathrm{iter}}$ is the number of complete iterations performed in a single run and the second factor is the number of independent restarts needed to reach the best-known solution with 99\% confidence. We report iterations rather than physical time because the present prototype is not optimized for execution speed, so wall-clock numbers would reflect the maturity of the setup rather than the advantages of the underlying architecture. Eq.~\ref{eq:ITS_def} depends on $N_{\mathrm{iter}}$ both explicitly and through $SR(N_{\mathrm{iter}})$, which increases with $N_{\mathrm{iter}}$ as short runs rarely succeed and therefore require many restarts, whereas long runs succeed often but are individually expensive. The product has a minimum, and quoting the ITS at an arbitrary $N_{\mathrm{iter}}$ is consequently ambiguous. Following common practice~\cite{im2026accelerating, de2025incorporating, sevenants2025requirements, sevenants2026reducing}, we therefore scan $N_{\mathrm{iter}}$ over the full range of iterations used and report the minimum. Instances with $SR = 0$ do not have a finite ITS and are marked with a cross ($\times$) in Table~\ref{tab:hyperparameters}. All ITS values are rounded to the nearest integer.
 
\pagebreak

\section{Feedback parallelization}


In the main text, it is shown that multiple problem instances can be programmed simultaneously onto the SLM. This is especially beneficial for smaller problems, as it increases the throughput of the setup by approximately an order of magnitude. However, the same figure also shows that this speed-up gradually saturates as the number of programmed problem instances increases. In this section, we investigate the underlying cause of this behavior and explore whether it can be mitigated. Note that there is a subtle difference from the problem-size scaling discussed in the main text. There, the number of spins per problem instance was varied, whereas here the problem size is fixed and we instead vary how many independent problem instances are programmed simultaneously. \newline

\begin{figure}[t!]
    \centering
    \includegraphics[width=0.85\linewidth]{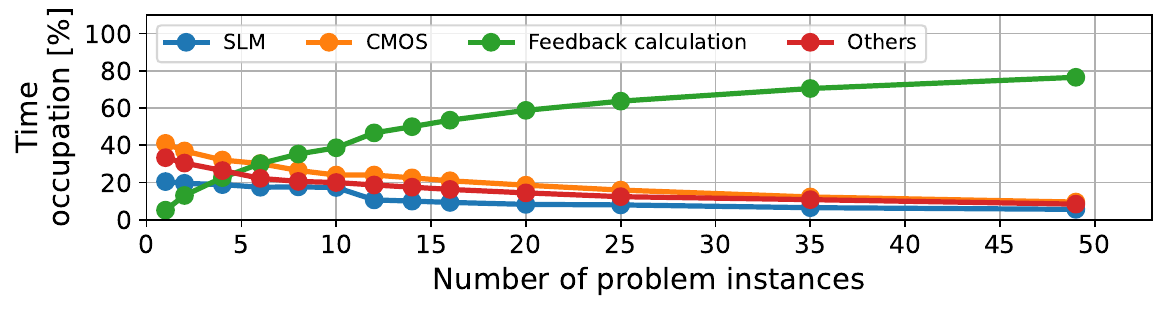}
    \caption{\textbf{Relative execution time distribution as a function of the number of problem instances}. Each color represents a different contributor.}
    \label{fig:time_occupation_1core_changing_processes}
\end{figure}

To understand the bottleneck, we analyze the distribution of the total runtime over the three main processing categories, together with the remaining runtime. These categories correspond to the time spent communicating with and updating the SLM, communicating with and transferring data to and from the CMOS camera, and calculating the feedback signal. The final category contains the remaining runtime, which is mainly associated with data transfer between function classes and file writing. Fig.~\ref{fig:time_occupation_1core_changing_processes} shows how this runtime distribution evolves as the number of programmed problem instances increases. Each category is represented by a different color, and each data point is obtained by averaging 10 runs of 50 iterations each. Note that the figure shows the percentage of the total runtime rather than the absolute execution times. \newline

\begin{figure}[b!]
    \centering
    \includegraphics[width=1\linewidth]{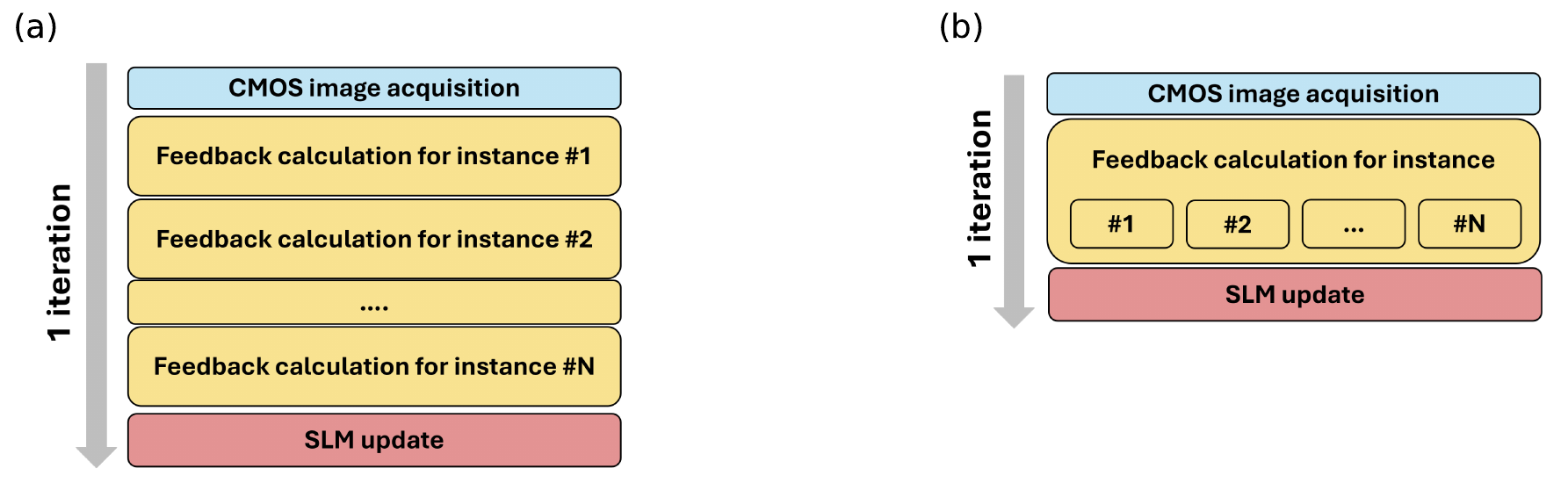}
    \caption{\textbf{Schematic illustration of the normal execution flow in our ASIM compared to the execution flow with parallelized digital processing.} \textbf{(a)} Normal execution flow, in which the parallel instances programmed on the SLM are processed sequentially on the CPU. \textbf{(b)} Parallelized execution flow, in which the feedback for all parallel instances programmed on the SLM is calculated simultaneously.}
    \label{fig:IM_execution_flow}
\end{figure}

As expected, the fraction of time spent calculating the feedback increases strongly with the number of problem instances, growing from approximately 5\% for a single problem instance to 76\% for 49 problem instances, thereby becoming by far the dominant contributor to the total runtime. The apparent decrease in the relative runtime of the other categories is a direct consequence of this increase. Their absolute execution times remain approximately constant, but their relative contribution decreases because the total runtime is increasingly dominated by the feedback calculations. \newline

Figure~\ref{fig:IM_execution_flow}(a) shows a schematic of the execution flow of our setup. First, a CMOS image is taken and each of the parallel instances programmed on the SLM is then processed digitally in sequence, after which the SLM is updated. Consequently, even if each individual feedback calculation takes only a short time, repeating this procedure $N$ times can become a bottleneck. Since each problem instance involves an identical workload of independent calculations, this feedback calculation time can be drastically reduced by exploiting its suitability for parallelization. In this scenario, each core processes its assigned problem instance independently, requiring minimal to no inter-core communication, which enables near-ideal scaling of the throughput up to the number of physically available cores. This principle is illustrated schematically in Fig.~\ref{fig:IM_execution_flow}(b), which shows how processing the problem instances in parallel reduces the feedback part of the execution flow from $N$ steps to a single step.\newline

\begin{figure}[b!]
    \centering
    \includegraphics[width=1\linewidth]{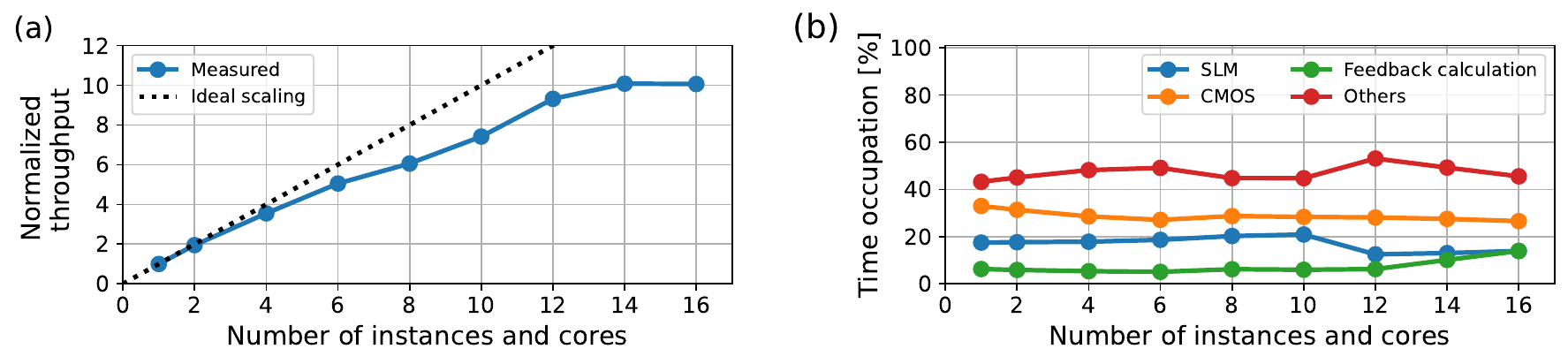}
    \caption{ \textbf{Number of cores equals the number of problem instances.} \textbf{(a)} Normalized throughput as a function of the number of cores and problem instances, where one problem instance is assigned to each core. The dotted line indicates ideal linear scaling, while the blue markers show the measured throughput normalized to the throughput of a single problem instance running on a single core. \textbf{(b)} Relative execution time distribution as the number of cores and problem instances increases under the same one-instance-per-core parallelization strategy. Each color represents a different component executed during each iteration. 
    }
    \label{fig:process_and_instance_scaling}
\end{figure}

To demonstrate this principle, we further improve the throughput of the ASIM implementation by distributing the feedback calculations across the cores of our Intel i7-10700 CPU (@2.90 GHz, 16 GB RAM), which provides 8 physical cores and, via hyperthreading, 16 logical threads. Specifically, each programmed problem instance on the SLM is assigned to a dedicated CPU core (or thread). The resulting normalized throughput is shown in Fig.~\ref{fig:process_and_instance_scaling}(a). The dotted line indicates the ideal scaling, while the blue data points show the measured normalized throughput. Initially, the throughput scales nearly ideally. However, beyond 12 problem instances running on 12 separate threads, the throughput begins to saturate. Nevertheless, an order-of-magnitude increase in throughput is already achieved using 14 problem instances on the same number of CPU threads, whereas approximately 35 problem instances were required to obtain the same speed-up when using only a single core. \newline

The corresponding runtime distribution is shown in Fig.~\ref{fig:process_and_instance_scaling}(b) with the same categories that we used before. The results in this figure confirm that the implemented parallelization is effective, as the time occupation of all categories remains approximately constant. Only beyond 14 problem instances do we observe an increase in the time required for the feedback calculations. 
The problem considered here, the 100-spin antiferromagnetic square lattice, is both relatively small and sparse. The latter allowed us to use the compressed sparse row (csr) method for a faster matrix-vector multiplication in the feedback-signal processing. For larger and fully connected problems, the time spent on the feedback calculations is expected to be significantly larger, suggesting that the achievable speed-up from parallelization could be even more pronounced in that case. Although we now used the same problem across all the parallel programmed instances, it would be possible to implement different problems at the same time. However, when problem instances of varying size or complexity are programmed simultaneously, the parallelization might also be less efficient, as the workload on each core will differ. Therefore, latency can start to show up as some cores remain idle while waiting for others to finish their calculations. 

\pagebreak

\section{Performance bottlenecks}

Each iteration of our setup consists of three stages: the SLM is updated, a CMOS image is acquired, and the feedback signal is computed from the average intensity measured by the CMOS camera. In the main text we report both the absolute and relative time spent in each stage, since together they reveal how the individual stages scale with problem size and how the resulting bottleneck shifts as the system grows. At small problem sizes ($N < 10^4$ spins), the SLM update is the dominant contribution to the iteration time. Because this time does not grow with $N$, increasing the problem size in this regime directly increases throughput. Beyond $N \geq 10^4$ spins, however, the feedback calculation becomes the limiting stage, since it involves a matrix-vector multiplication (MVM) whose cost scales as $\sim N^2$. Note that the timings reported correspond to the full, unsparsified MVM. Each of the three main components (CMOS, SLM, and feedback) is not a single, monolithic operation, but instead comprises several distinct steps. In the following subsections, we analyze each component in more detail to identify its specific dominant bottlenecks and determine where the largest speed improvements can be achieved. Since the analysis will be performed on their relative contributions, we only look at the relative time distribution.

\subsection{CMOS time distribution}

\begin{figure}[b!]
    \centering
    \includegraphics[width=1.0\linewidth]{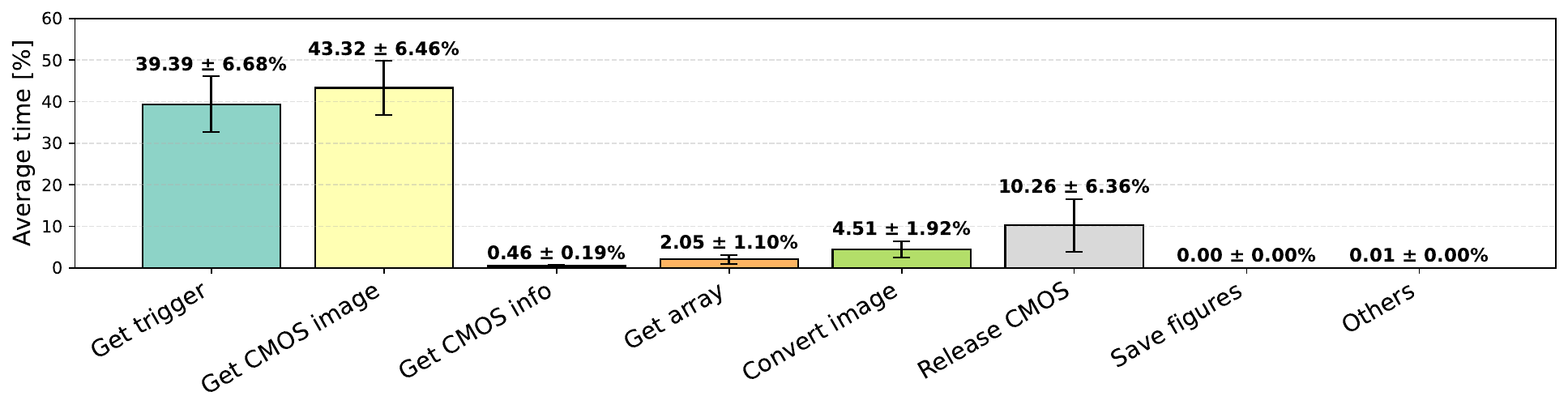}
    \caption{\textbf{Time distribution of processing an image with the CMOS camera.} Each bar represents the mean value for each distinct part of the process while the error bars represent the standard deviation. The mean and standard values are calculated from 500 iterations and are written on top of the bars.}
    \label{fig:CMOS_time_distribution}
\end{figure}

We start by further analyzing the individual steps contributing to the total CMOS image processing time. This time does not contain only 'taking the picture' but also triggering the CMOS camera to acquire an image, transferring the CMOS image to the desktop, retrieving information from the CMOS image, converting the image into a \texttt{NumPy} 16-bit array, converting the 16-bit image data back to its original 12-bit representation, free the CMOS buffer, checking whether the image should be saved (without actually performing the save operation in this measurement), and other processing steps. The relative time contribution of each of these steps is shown in Fig.~\ref{fig:CMOS_time_distribution}. The bars represent the average value, calculated from the 500 iterations and the error bars represent the associated standard deviations. The exact values are also written on top of the bar.  \newline

The figure shows that triggering the CMOS camera and transferring the image to the computer are the dominant contributors to the acquisition time. The remaining processing steps together account for around 15\% of the total time. These two stages therefore represent the main bottlenecks in the CMOS stage of the implementation. The camera trigger process could be optimized by implementing an analog trigger that is activated directly when the SLM update is complete. This would avoid unnecessary waiting time and could reduce the trigger delay from the millisecond range to the microsecond range. Since CMOS processing accounts for approximately 30\% of the total execution time, and triggering represents approximately 40\% of the CMOS processing time, eliminating or significantly reducing the triggering overhead could reduce the overall execution time by approximately 12\%. The image transfer time could potentially be reduced by improving the data transmission between the CMOS sensor and the computer. However, due to the relatively small size of the image ($\approx$ 6MB) compared to the already sufficient bandwidth of the CMOS camera’s USB 3.0 interface (625 MB/s), increasing the bandwidth is not expected to have a significant impact. A more detailed investigation is required to determine where further time savings can be achieved in this category. However, it is likely that the USB protocol introduces additional latency during image transfer, which could contribute to the relatively high time consumption. Another possible contributor is the sensor readout time, where the analog signal from the CMOS sensor is converted into digital data before being transferred to the computer. \newline


\subsection{SLM time distribution}

Next, we analyze the time distribution associated with updating the SLM. Also this process involves more than simply refreshing the displayed phase pattern. It includes generating the desired phase pattern, sending the command to the SLM, and waiting for the LCs within the pixels to reach their new state. The latter step is required because there is a finite delay between refreshing the SLM display (i.e., the refresh rate) and the actual response time of the liquid crystals. Since the SLM only provides feedback once the display refresh has been completed, an additional waiting time is introduced to ensure that the desired phase pattern has fully settled. In our setup, this waiting time was set to 100~ms, which corresponds to approximately half of the time required for the liquid crystals of this SLM model to rotate by 180$^\circ$, as specified by the manufacture. \newline

\begin{figure}[t]
    \centering
    \includegraphics[width=1.0\linewidth]{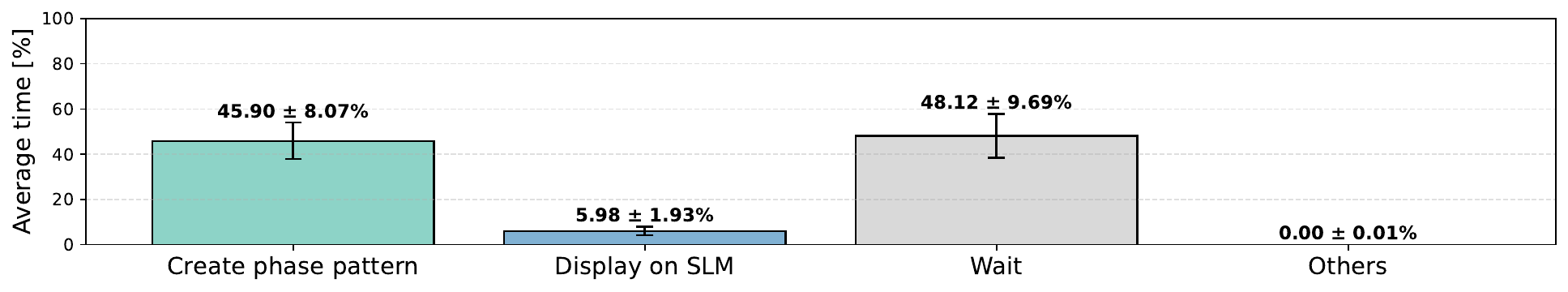}
    \caption{\textbf{Time distribution of putting an phase pattern on the SLM.} Each bar represents the mean value for each distinct part of the process while the error bars represent the standard deviation. The mean and standard values are calculated from 500 iterations and are written on top of the bars.}
    \label{fig:SLM_time_distribution}
\end{figure}

Figure~\ref{fig:SLM_time_distribution} shows the relative time distribution of the four different components involved in updating the SLM. The bars represent the average values calculated from 500 iterations, while the error bars indicate the corresponding standard deviations. The exact values are also shown above each bar. Two components clearly dominate the total time required for an SLM update. The first is the generation of the phase pattern, which is currently performed on the host computer using \texttt{NumPy}. This step could potentially be accelerated by using a faster or more optimized implementation. The second major contribution is the waiting time. As discussed above, this delay is introduced to ensure that the desired phase pattern has fully settled before the next iteration begins. The current waiting time of 100~ms was deliberately chosen as a conservative value. However, during most iterations the phase pattern changes only slightly, meaning that the liquid crystals typically undergo a rather small rotation. Consequently, the required settling time is expected to be shorter in most cases, suggesting that this waiting period could be further optimized to reduce idle time in future implementations. \newline


\subsection{Feedback time distribution}

\begin{figure}[b!]
    \centering
    \includegraphics[width=1.0\linewidth]{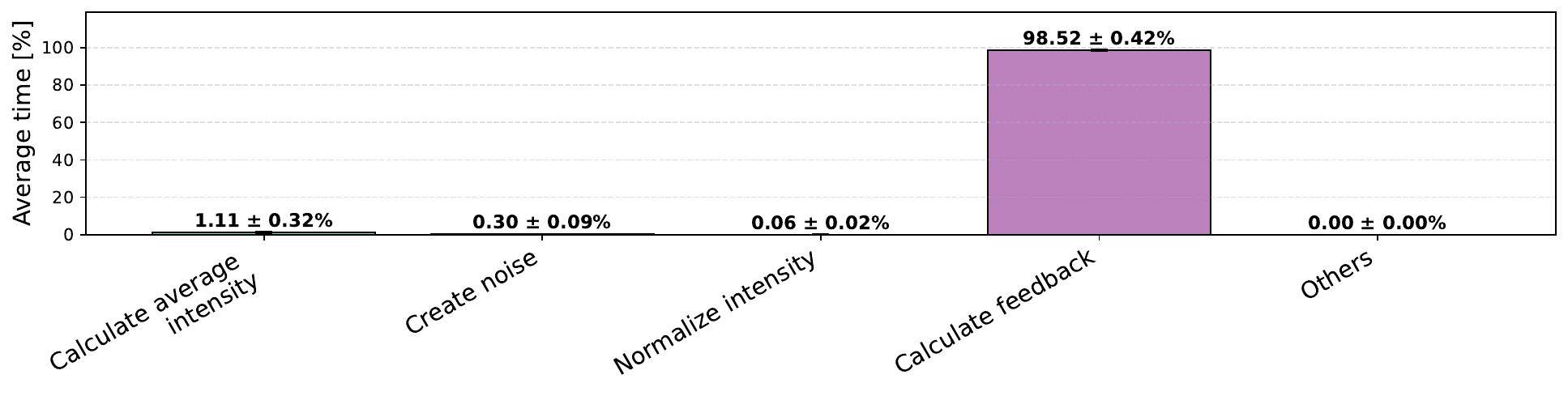}
    \caption{\textbf{Time distribution of calculating the feedback signal for a 25{,}600-spin problem, when performing the full matrix-vector multiplication.} Each bar represents the mean value for each distinct part of the process, while the error bars represent the standard deviation. The mean and standard deviation are calculated from 500 iterations and are indicated on top of the bars.}
    \label{fig:FEEDBACK_time_distribution_full}
\end{figure}

Finally, we analyze how the computation time within the feedback stage is distributed. This stage consists of several operations: calculating the average intensity from the cluster of CMOS pixels representing a single spin, generating and applying digital noise to the spin amplitudes, normalizing the spin amplitudes, and computing the feedback signal itself. Since the main bottleneck of the feedback stage arises at larger problem sizes, we use the largest problem size for which a full MVM could still be run, namely 25{,}600 spins. The relative time distribution of these operations for this problem size is shown in Fig.~\ref{fig:FEEDBACK_time_distribution_full}. This figure shows that calculating the feedback dominates the computational cost, accounting for more than 98\% of the total time spent in the feedback stage, consistent with what is expected when performing the full MVM. This also implies that offloading the MVM to the optical domain could strongly reduce the largest bottleneck of the feedback stage, and, at these problem sizes, of the total iteration time. \newline

In the meantime, this cost can already be reduced using sparsification methods. In particular, we used the \textit{csr} method to calculate the feedback signal for the largest antiferromagnetic square-lattice problems ($\geq 10^4$ spins). Sparsification is effective for this benchmark problem because, even though the number of couplings per spin remains fixed as the system grows,
\begin{equation}
    \frac{E}{N} = 2,
\end{equation}
where $E$ is the number of edges and $N$ the number of spins, the coupling matrix itself becomes increasingly sparse as its density scales as
\begin{equation}
    \rho = \frac{E}{N(N-1)} \approx \frac{2}{N},
\end{equation}
which for $N = 25{,}600$ corresponds to $\rho \approx 7.81\times10^{-5}$, i.e.\ roughly $0.0078\%$ non-zero off-diagonal entries. Sparsification then removes the cost of also processing the fraction of zero entries that a dense $N\times N$ representation would otherwise carry, reducing the MVM from $O(N^2)$ to $O(N)$. \newline

\begin{figure}[t!]
    \centering
    \includegraphics[width=1.0\linewidth]{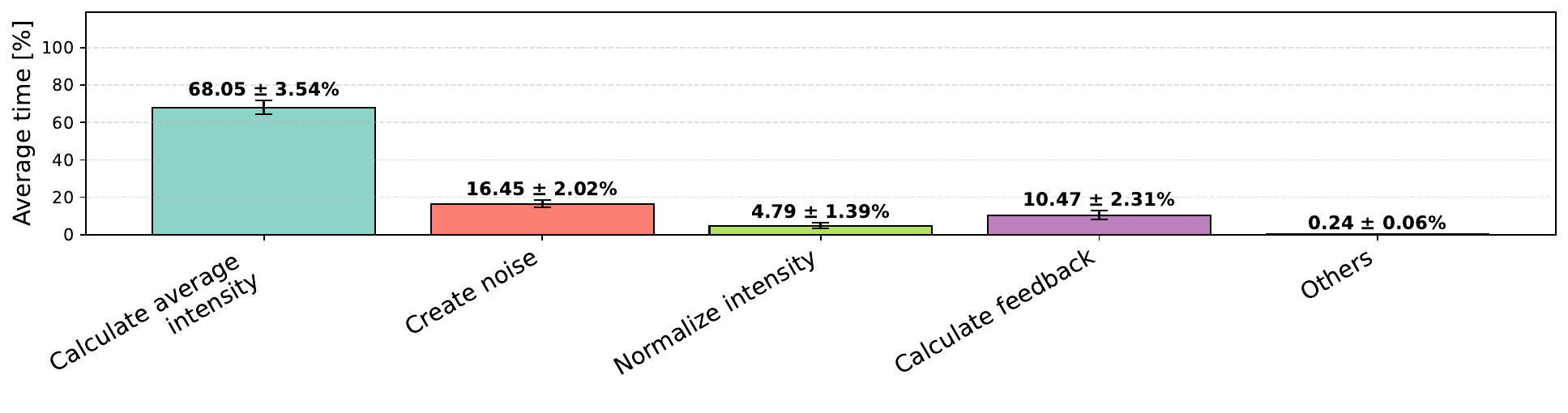}
    \caption{\textbf{Time distribution of calculating the feedback signal for a 25{,}600 spin problem, using the csr method.} Each bar represents the mean value for each distinct part of the process while the error bars represent the standard deviation. The mean and standard values are calculated from 500 iterations and are written on top of the bars.}
    \label{fig:FEEDBACK_time_distribution_sparse}
\end{figure}

To assess the impact of sparsification on the time distribution, we again examine the relative time distribution of the same operations, for the same 25{,}600-spin case, shown in Fig.~\ref{fig:FEEDBACK_time_distribution_sparse}. Sparsification has a strong effect, reducing the share of time spent calculating the feedback from 98\% to about 10\%, corresponding to a drop from roughly 333\,ms to 0.24\,ms on average. As a result, calculating the feedback is no longer the main bottleneck. Instead, computing the average intensity corresponding to each spin amplitude becomes dominant. This is because each spin amplitude is not read out from a single CMOS pixel, but is instead reconstructed from a cluster of several CMOS pixels. \newline

Consequently, for each spin the corresponding pixel values must first be gathered from the 2D CMOS pixel array and copied into a temporary array before their mean is computed. The cost of this operation scales with the number of pixels retrieved rather than with the arithmetic of the averaging itself, and since the number of pixels per spin is roughly constant, we expect this cost to grow linearly with the number of spins. However, the average-intensity calculation, which accounts for 68\% of the feedback time in Fig.~\ref{fig:FEEDBACK_time_distribution_sparse}, corresponds to only 1.63\,ms on average, still much smaller than the 160\,ms that the SLM steering contributes to each iteration at this problem size.


\newpage

\section{Power consumption}

A significant part of the motivation behind application-specific, novel computing paradigms is that traditional digital computers spend considerable time and energy
(the two being closely related) executing a given task. Consequently, for novel hardware implementations it is not only important to report performance in terms of
success rate or time- or iteration-to-solution, but also the energy consumed during computation. \newline

However, energy consumption is considerably harder to quantify than computation time. Unlike time, which can be measured directly and consistently for a given operation, power draw is not constant. It depends on the instantaneous workload of each component, can fluctuate over the course of a measurement (e.g. due to thermal throttling or duty-cycle variations), and, for shared components such as the CPU, reflects the load of the entire system rather than the feedback calculation alone. Furthermore, dedicated power meters are not available for every individual component, so estimates must in some cases rely on manufacturer specifications rather than direct measurement. Nevertheless, given its importance, we provide a more detailed breakdown of the energy consumption per component, mainly in an attempt to identify the most energy-intensive part of the system. Tab.~\ref{tab:energy_consumption} lists the power consumption of the main hardware components of the setup: the spatial light modulator (SLM), the CMOS sensor, the CPU used to compute the feedback signal, the LED used as the light source and the power driver that generates the energy of the LED. \newline

As seen from Tab.~\ref{tab:energy_consumption}, we expect that the total power consumption of the system is approximately 100~W, with the CPU and the SLM together accounting for the large majority of this value. This indicates that, in terms of energy, improvements to the feedback-calculation hardware and the SLM driving electronics would have the largest impact on the overall energy efficiency of the system. \newline 

We expect the total energy consumption of our setup to scale only weakly with problem size. The LED, CMOS, and SLM operate at approximately constant power and their operating time per iteration is largely independent of problem size, so their contribution to the total energy consumption remains roughly constant regardless of the problem size. The CPU is the only component whose energy consumption is expected to scale with problem size, since the feedback-calculation time itself increases with problem size. This weak overall scaling is not unique to our implementation and is expected in other measurement--feedback schemes that similarly rely on a digital component to compute the feedback signal. However, we do expect a big contrast with time-multiplexed architectures, in which spins are encoded sequentially, so that the total encoding time, and therefore the energy consumed by every component of the setup, scales directly with problem size.

\begin{table}[h]
    \centering
    \caption{Power consumption of the ASIM, broken down by component.}
    \label{tab:energy_consumption}
    \begin{tabularx}{\textwidth}{|l|c|X|}
        \hline
        \textbf{Component} & \textbf{Power} & \textbf{Notes} \\
        \hline
        SLM  & 25.00~W & Maximum power rating on the power adapter \\
        \hline
        LED  & 1.04~W  & Calculated from the measured current and voltage consumption \\
        \hline
        CMOS & 4.50~W  & Typical value, given by manufacturer \\
        \hline
        CPU  & 65.00~W & Typical value, given by manufacturer \\
        \hline
        \textbf{Total} & \textbf{95.54~W} & \\
        \hline
    \end{tabularx}
\end{table}

\printbibliography[
  heading=bibliography,
  title={References}
]